\documentclass[a4paper,fleqn]{cas-dc}

\usepackage[authoryear,longnamesfirst]{natbib}
\usepackage{amsmath}
\usepackage{amssymb}
\usepackage{algorithm}
\usepackage{algorithmic}
\usepackage{booktabs}
\usepackage{array}
\usepackage{float}
\usepackage{multirow}
\usepackage{xcolor}
\usepackage{subcaption}
\usepackage{pifont}
\usepackage{graphicx}
\usepackage{cuted}

\floatname{algorithm}{Procedure}

\newenvironment{nonfloattable}
  {\par\medskip\noindent\begin{minipage}{\columnwidth}\captionsetup{type=table}}
  {\end{minipage}\par\medskip}

\def\tsc#1{\csdef{#1}{\textsc{\lowercase{#1}}\xspace}}
\tsc{WGM}
\tsc{QE}

\begin{document}
\let\WriteBookmarks\relax

\shorttitle{Latent Declarative Representations for Repository Migration}

\shortauthors{S. Surana et al.}

\title [mode = title]{Identifying Latent Declarative Representations of Code for Assisting Repository Migration}



%

\author[1]{Shraddha Surana}[orcid=0000-0002-3009-3178]
\cormark[1]
\ead{p20220031@goa.bits-pilani.ac.in}

\author[1]{Ashwin Srinivasan}
\ead{ashwin@goa.bits-pilani.ac.in}

\author[2]{Michael Bain}
\ead{m.bain@unsw.edu.au}

\affiliation[1]{organization={Department of Computer Science and Information Systems, BITS Pilani, K. K. Birla Goa Campus},
                addressline={NH 17B, Zuarinagar},
                postcode={403726},
                city={Goa},
                country={India}}

\affiliation[2]{organization={School of Computer Science and Engineering, UNSW Sydney},
                city={Sydney},
                state={NSW},
                postcode={2052},
                country={Australia}}

\cortext[1]{Corresponding author}



\begin{abstract}
Legacy software repositories embed decades of domain knowledge in
undocumented code, making understanding and modernization difficult.
We treat a program as the implementation of an unobserved, declarative
description of its computation and investigate whether making this
\emph{latent declarative representation} explicit improves
repository-scale porting. ADFD-Migrate approximates the latent
representation with an annotated data-flow diagram (ADFD) of processes,
data stores, external entities, flows, and behavioral contracts. An LLM
infers the source ADFD from bounded repository context, guided by static-analysis coverage checks. Dependency-aware
chunking orders bounded process groups for target-language generation.
Differences between the source ADFD and a statically recovered target ADFD
then guide regeneration.
We evaluate ADFD-Migrate on {\tt f2x50}, a new benchmark of 50~Fortran repositories spanning
1.5k--1.6M lines of code and three complexity tiers,
and assess the resulting ports along two dimensions: porting soundness,
measured by source-oracle behavioral agreement, and porting completeness,
measured by a composite migration outcome index.
Against 382 curated Fortran-oracle probes, the generated Python passes
327 (85.6\%), with 40 repositories passing every attempted probe.
ADFD-Migrate exposes all 382 planned behaviors as runnable targets,
compared with 99 and 98 for direct and repository-context translation
and 69 and 30 for the static-profile and dependency-chunking ablations.
It also achieves a 93.1\% mean migration outcome index and a 17--59
percentage-point outcome-index advantage over direct translation on 47 repositories.
These results suggest that an inspectable semantic bottleneck can improve
the coverage and integration of repository-scale migration while enabling
lower-cost generation for many repositories.
\end{abstract}



\maketitle

\section*{Glossary}
\begin{description}
\item[ADFD] Annotated data-flow diagram: the intermediate representation
of processes, data stores, external entities, flows, and process contracts.
\item[Behavioral probe] A source-oracle test that compares Fortran and
generated Python behavior on the same input.
\item[Latent declarative representation] The unobserved,
language-independent specification of the computation implemented by a
program. An inferred ADFD is an explicit, inspectable approximation of
this representation.
\item[Migration outcome index] The composite measure of porting
completeness, combining the generated migration's semantic alignment,
executability, code quality, and implementation rate.
\item[SCC] Strongly connected component in the ADFD process-dependency graph.
\end{description}

\section{Introduction}\label{sec:intro}

Legacy software systems, often written in languages such as Fortran,
COBOL, or early C, represent decades of accumulated domain knowledge
embedded in code~\cite{vonmayrhauser1995program}.  When the original
developers are unavailable and documentation is sparse, understanding
these systems becomes a major challenge for software maintenance and
modernization efforts~\cite{comella2000survey}.  Scientific computing is
particularly affected: critical numerical libraries (LAPACK, MINPACK, WRF)
remain in active use but are locked in Fortran, limiting integration with
modern data science ecosystems built around Python.

Recent advances in large language models~(LLMs) have demonstrated
impressive capabilities in code understanding and
generation~\cite{chen2021codex, roziere2023codellama}.  However, direct
LLM-based translation of legacy code, i.e., asking the model to translate an
entire repository directly, suffers from several limitations:
(1)~loss of cross-module structure, (2)~hallucinated APIs and
dependencies, (3)~no mechanism for detecting or correcting errors, and
(4)~no intermediate representation that a human can inspect and
validate~\cite{pan2024lost}.

We start from the premise that a program is one procedural implementation
of a more declarative description of its computation. That description is
\emph{latent}: it is not directly observed as a separate artifact in the
repository. We use an Annotated Data Flow Diagram~(ADFD) as an explicit
hypothesis about this latent declarative representation. The ADFD records
what the computation must do through processes, data stores, external
entities, flows, and behavioral contracts, without committing to the
source language's implementation details.

\begin{figure*}[t]
\centering
\includegraphics[width=0.98\textwidth]{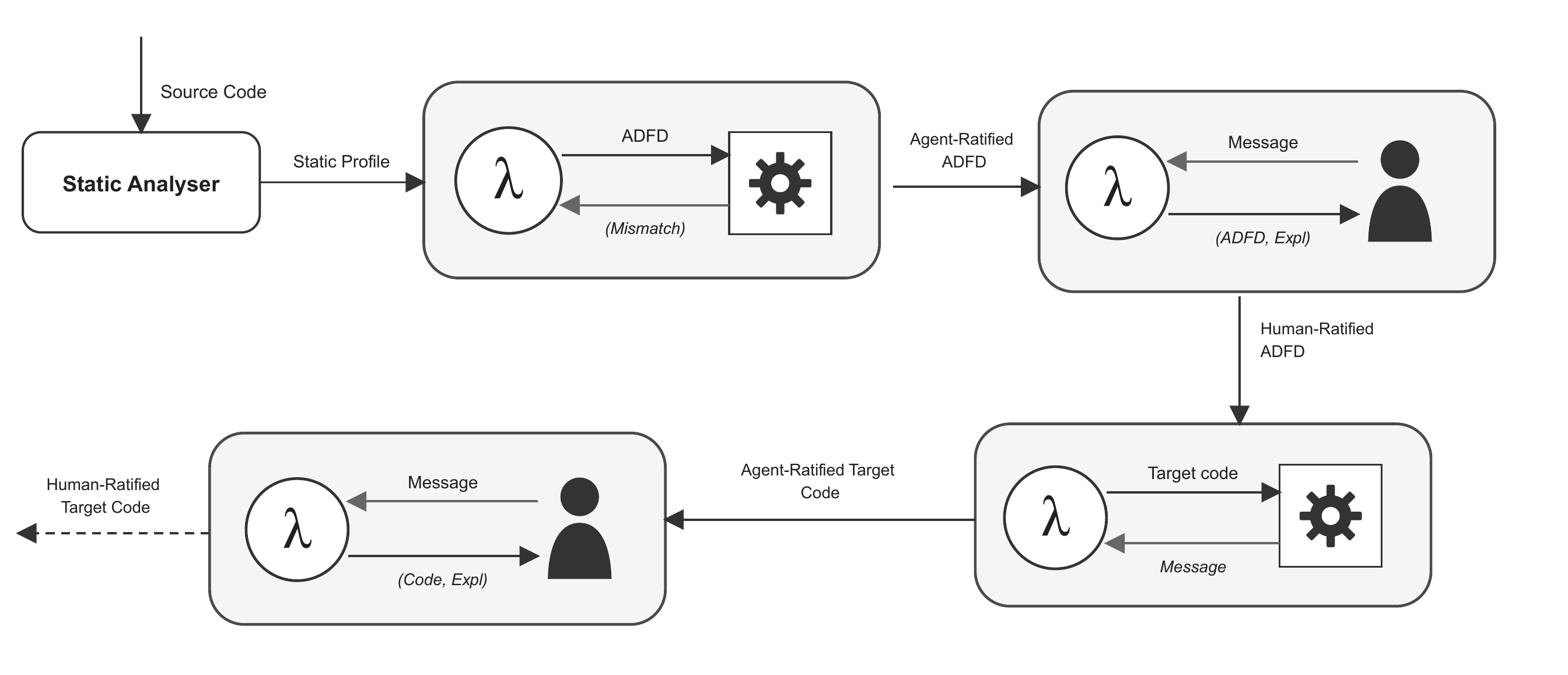}
\caption{ADFD-mediated porting workflow. The $\lambda$ and gear symbols
denote LLM and deterministic agents. The first diagnostic loop refines a
source ADFD; after optional human ratification, the second refines target
code. Experiments automate both loops; human exchanges indicate where
  experts can optionally inspect and correct the intermediate ADFD and target code.}
\label{fig:pipeline}
\end{figure*}

Figure~\ref{fig:pipeline} shows the ADFD-mediated porting workflow.
Static analysis first produces a static source profile. An LLM proposes an ADFD,
and a deterministic agent returns mismatches until the ADFD is
agent-ratified; a human expert may then correct it before ratification.
The ratified ADFD becomes the specification for target-code generation, followed by an
analogous diagnostic loop and expert review. Thus, the explicit latent
representation is checked before being reused to construct the target.

ADFD-Migrate operationalizes the machine-facing parts of this workflow:
static-analysis coverage supplies the mismatch signal during source-ADFD
construction, while dependency-aware generation and source--target ADFD
comparison provide the target-side feedback loop. The experiments run the
two LLM--software-agent interactions automatically, so the
human-ratification interactions are inspectable extension points rather
than an evaluated treatment. Appendix~\ref{app:full-architecture} maps
this workflow to the complete automated system architecture.

A central contribution is the \emph{dependency-aware chunking} strategy used to organize target-code generation. Tarjan's SCC
detection~\cite{tarjan1972depth} and Kahn's topological
sorting~\cite{kahn1962topological} partition large ADFDs into bounded,
dependency-ordered process groups. Each group is generated with upstream
interface summaries, keeping calls tractable while preserving cross-module
references.

By reducing per-call complexity, chunking makes smaller open-source models
usable for the computationally intensive generation. Low- and medium-tier generation ran on Qwen-3-Coder
(30B locally via Ollama, 480B cloud-hosted) and incurred no metered
per-call charges; complex-tier generation and the evaluation-only judge
used Anthropic models. Appendix~\ref{sec:cost-analysis} states the
cost-accounting boundary and measured charges.

This paper makes the following contributions:
\begin{enumerate}
  \item \begin{sloppypar}An ADFD-mediated migration framework that treats
        the declarative specification of the computation implemented by a
        legacy repository as a latent representation and makes an
        explicit, inspectable approximation of it as an ADFD for
        cross-language code porting with an iterative source ADFD construction loop
        driven by static coverage scoring
        (Section~\ref{sec:source-adfd}). This is distinct from call-graph
        decomposition~\cite{ibrahimzada2025alphatrans}, intermediate
        programming languages~\cite{macedo2025intertrans}, and compiler
        IRs~\cite{szafraniec2023code}.\end{sloppypar}
  \item A \emph{dependency-aware chunking} algorithm based on Tarjan's
        SCC detection and topological sorting that partitions ADFD process
        \emph{dependency graphs} into bounded, topologically ordered
        groups while preserving referential integrity
        (Section~\ref{sec:chunking}).
  \item {\tt f2x50}, a curated benchmark of 50
        open-source Fortran repositories across three complexity tiers
        with converted Python repositories, intermediate ADFDs, and
        evaluation artifacts (Section~\ref{sec:benchmark}).
  \item A correctness-first empirical evaluation across all 50
        repositories, comparing ADFD-Migrate against four non-ADFD
        baselines and ablations: direct file-by-file translation, repository-context
        direct translation, static-profile ablation, and
        dependency-chunking ablation. ADFD-mediated decomposition reaches 85.6\% behavioral
        agreement over all planned probes against 21.7\% for both direct
        and repository-context direct translation and 11.0\% and 3.7\% for
        the two ablations, and a
        +17 to +59~pp outcome-index advantage, with the
        largest gains on complex repositories and with reduced or
        eliminated API cost (Section~\ref{sec:results}).
\end{enumerate}

Section~\ref{sec:related} positions the approach against existing
migration tools and LLM translation methods.
Section~\ref{sec:approach} defines the ADFD and the porting procedures,
Section~\ref{sec:eval-design} states the hypothesis and experimental
design, Section~\ref{sec:results} reports results along the two dimensions
of porting effectiveness, and Section~\ref{sec:discussion} discusses the
method and threats to validity.

\section{Background and Related Work}\label{sec:related}

\paragraph{Legacy Code Modernization}

Table~\ref{tab:tool-comparison} positions our approach relative to
existing Fortran migration tools.  \texttt{f2py}~\cite{peterson2009f2py}
generates Python bindings that call compiled Fortran via C extensions. It does not produce readable Python.
\texttt{fable}~\cite{grosse2012fable} performs rule-based
Fortran-to-C++ translation but the output retains Fortran idioms
requiring manual post-editing.
\texttt{LFortran}~\cite{lfortran2023} compiles to LLVM IR with an
experimental AST-to-Python transpiler limited to a subset of Fortran~2018. Our approach translates through a \emph{semantic}
intermediate representation that captures \emph{what} the code computes and
not simply \emph{how} it is expressed.

\begin{table}[t]
\caption{Comparison of Fortran migration approaches.}\label{tab:tool-comparison}
\centering
\small
\resizebox{\columnwidth}{!}{%
\begin{tabular}{@{}lcccc@{}}
\toprule
\textbf{Property} & \textbf{f2py} & \textbf{fable} & \textbf{LFortran} & \textbf{Ours} \\
\midrule
Target language      & Py (bind)    & C++       & Py (AST)   & Python \\
Readable output      & \ding{55}    & Partial   & \ding{55}  & \ding{51} \\
Idiomatic target     & \ding{55}    & \ding{55} & \ding{55}  & \ding{51} \\
Preserves structure  & N/A          & \ding{55} & \ding{55}  & \ding{51} \\
Inspectable IR       & \ding{55}    & \ding{55} & LLVM IR    & ADFD \\
Human checkpoint     & \ding{55}    & \ding{55} & \ding{55}  & \ding{51} \\
\bottomrule
\end{tabular}
}
\end{table}

\paragraph{LLM-Based Code Translation}
\begin{sloppypar}
TransCoder~\cite{roziere2020transcoder} uses unsupervised machine
translation between programming languages.\end{sloppypar}  General-purpose LLMs show
strong performance on code translation benchmarks~\cite{yang2024exploring,
yin2024rectifier} but typically operate at the file or function level.
Code Distillation (CoDist)~\cite{codist2023} uses a language-agnostic
distilled-code pivot, improving over TransCoder-ST by 12.7\% on average in
snippet benchmarks. Unlike our ADFD, it creates parallel training corpora
without addressing repository-wide dependency ordering, interface
propagation, or human-inspectable checkpoints. Recent work
on long code blocks shows that context management remains central for
large translation tasks~\cite{chakaravarthy2026longblocks}.
Saha et al.~\cite{SAHA2026112964} investigate natural language
specifications as an intermediate representation across five languages,
finding inconsistent gains alone but improvements for some language pairs
when combined with source code. Our work instead studies how an explicit
ADFD representation can organize repository-scale
migration, where dependency structure, interfaces, import paths, and
generated artifacts must remain coherent across a 
repository.

\paragraph{Repository-level translation} This is an active frontier.
AlphaTrans~\cite{ibrahimzada2025alphatrans} decomposes Java repositories
into program fragments using static analysis, translates them in reverse
call order, and validates translated fragments using the source project's
tests. It reports 96.40\% syntactically correct fragments and validates
runtime behavior and functional correctness for 27.03\% and 25.14\% of
application-method fragments. These results are not directly comparable:
AlphaTrans studies ten Java-to-Python projects using project test suites,
whereas we use uniform source-oracle probes for Fortran-to-Python. No shared
benchmark supports a controlled head-to-head test.
InterTrans~\cite{macedo2025intertrans} routes through intermediate
programming languages to improve code-translation accuracy.
CodePlan~\cite{bairi2024codeplan} uses planning over dependency graphs.
SemanticForge~\cite{zhang2025semanticforge} constructs knowledge graphs for
repository-level generation.

Our approach differs in two ways: (1)~we translate through a
\emph{semantic} intermediate representation that captures computational
intent rather than syntactic structure, and (2)~the ADFD is designed for
human inspectability.  AlphaTrans decomposes by call graph (a syntactic
artifact); we decompose by \emph{data flow} (a semantic artifact).

\paragraph{Data Flow Diagrams (DFDs) in Software Engineering}

DFDs have a long history in structured systems
analysis~\cite{demarco1979structured, yourdon1989modern}.
Recent work has revived interest in DFDs as intermediate representations
for LLM-driven code generation.  Prior work on interactive program
synthesis~\cite{surana2026iprog} uses DFDs as specifications for
human-in-the-loop validation.  The present work extends this from
interactive code \emph{construction} to automated code \emph{porting},
using the DFD as a language-agnostic intermediate representation rather than
just a specification to be implemented.

\paragraph{LLM-Based Coding Agents}

\begin{sloppypar}
Autonomous coding agents such as SWE-Agent~\cite{yang2024sweagent},
Devin~\cite{cognition2024devin},
AutoCodeRover~\cite{zhang2024autocoderover} and
OpenHands~\cite{wang2024openhands} couple LLMs with tool use for
software engineering tasks.  AgenticTyper~\cite{pohle2026agentictyper}
automates JavaScript-to-TypeScript modernization;
FreshBrew~\cite{may2026freshbrew} benchmarks agents on Java~8-to-17
migration.
\end{sloppypar}

For repository-scale \emph{cross-language} porting, these agents usually
depend on frontier or task-specific models (AgenticTyper degrades with
smaller models), expose no inspectable intermediate artifact, and require
replaying long trajectories to debug failures. ADFD-Migrate instead uses
bounded chunks that localize errors, support a locally hosted 30B~model for
many repositories, and let experts inspect the ADFD before generation.

\section{Our Approach}\label{sec:approach}

We distinguish the latent declarative representation from the explicit
artifact used to approximate it. A repository is an observed
implementation, whereas its declarative specification of what is
computed is unobserved. The source ADFD is an inferred, explicit
approximation of that latent declarative representation. It allows different implementations to be mediated by a
shared representation while retaining the software-engineering benefits
of a declarative DFD: language independence, visual inspectability, and
composition from guarded processes.

The ADFD also serves as an \emph{explicit semantic bottleneck} because
target-code generation is conditioned on the bounded ADFD rather than
directly on the source repository. Information needed for generation is
organized into processes, data stores, entities, flows, and behavioral
contracts. The resulting artifact can be checked
against static analysis, reviewed by a developer, partitioned for
repository-scale generation, and compared across source and target
languages. We assume ADFDs are sufficiently expressive to approximate the
computational structure needed for migration, and specify the approach
using two definitions.

\noindent\textbf{Definition~1 (Annotated Data Flow Diagram).}
An ADFD is a tuple $(P, D, E, F, \sigma)$ where
$P$ is a set of \emph{processes},
$D$ a set of \emph{data stores},
$E$ a set of \emph{external entities},
$F \subseteq (P \cup D \cup E) \times (P \cup D \cup E)$ a set of
directed \emph{flows} (i.e., ordered pairs denoting who sends data to
whom), and
$\sigma\colon P \to (\textsf{Spec}, \textsf{Pre}, \textsf{Post},
\textsf{In}, \textsf{Out})$ assigns each process a specification,
pre/post-conditions, and typed I/O. We denote the set of
all possible ADFDs as $\mathcal{A}$.

\noindent\textbf{Definition~2 (ADFD-Mediated Porting of Repositories).}
Let $S$ denote a source implementation language and $T$ a target
implementation language. Let $\mathcal{R}_S$ and $\mathcal{R}_T$ denote
the set of all possible repositories in $S$ and $T$ respectively. We
use an ADFD as an explicit approximation of a repository's latent
declarative representation, and let $\mathcal{A}$ denote the space
of ADFD representations. Let
$\mathit{Construct}_S:\mathcal{R}_S \rightarrow \mathcal{A}$ and
$\mathit{Construct}_T:\mathcal{R}_T \rightarrow \mathcal{A}$ denote functions
that construct ADFDs from source- and target-language repositories,
respectively, and let
$\mathit{Generate}_T:\mathcal{A} \rightarrow \mathcal{R}_T$ denote a function
that generates a target repository from an ADFD. The porting function is
$\mathit{Port}_{S\rightarrow T} = \mathit{Generate}_T \circ
\mathit{Construct}_S$. After generation, source--target ADFD alignment compares
$A_S = \mathit{Construct}_S(R_S)$ with
$A_T = \mathit{Construct}_T(R_T)$. This separate operation evaluates and
refines the generated repository.

\subsection{Implementation}
We implement $\mathit{Construct}_S$, $\mathit{Construct}_T$, and
$\mathit{Generate}_T$ with ADFDs as the intermediate representation, but
the two construction functions are not identical. The source ADFD is
inferred by an LLM and checked against a Fortran static-analysis profile.
The target ADFD is extracted by Python static analysis, with optional LLM
enrichment of flow labels. Target-code refinement then uses structural
alignment between the source and target ADFDs. A separate LLM judge
assesses semantic alignment only for evaluation.

\subsubsection{Constructing and Checking the Source ADFD}\label{sec:source-adfd}

Procedure~\ref{proc:construct} constructs a source ADFD through iterative
refinement of the LLM's output. Static analysis of the source code
acts as a \emph{repository profile}, containing module structure, call graphs,
function/subroutine signatures, and file-to-definition mappings. The construction procedure
uses this repository profile together with relevant source-code excerpts to
generate an ADFD for the repository.
``Goodness'' is computed using a \emph{static coverage scoring} function that compares
the ADFD against the profile to identify \emph{gaps} (source files and
definitions not represented in the ADFD).  Targeted
directives describing the gaps are then injected into a prompt for the next
invocation of the LLM.

\begin{algorithm}[t]
\caption{\textsc{ConstructADFD}}\label{proc:construct}
\begin{algorithmic}[1]
\STATE \textbf{Input:} $R_S$: a source repository;\;
         $\lambda$: an LLM;\;
         $k$: a bound on iteration;\;
         $\tau$: a coverage threshold
\STATE \textbf{Output:}  $A$: an Annotated DFD
\STATE $\Sigma \gets \textsc{StaticAnalysis}(R_S)$
\STATE $A_{\text{best}} \gets \emptyset$;\;
       $s_{\text{best}} \gets 0$;\;
       $\mathit{dir} \gets \emptyset$
\FOR{$i = 1$ \TO $k$}
  \STATE $A_i \gets
        \textsc{InferADFD}(R_S,\lambda,\Sigma,\mathit{dir})$
  \STATE $(s_i, \mathit{gaps}_i) \gets
         \textsc{CoverageScore}(\Sigma, A_i)$
  \IF{$s_i > s_{\text{best}}$}
    \STATE $A_{\text{best}} \gets A_i$;\;
           $s_{\text{best}} \gets s_i$
  \ENDIF
  \IF{$s_{\text{best}} \ge \tau$}
    \STATE \textbf{break}
  \ENDIF
  \STATE $\mathit{dir} \gets
         \textsc{GapDirectives}(\mathit{gaps}_i)$
\ENDFOR
\RETURN $A_{\text{best}}$
\end{algorithmic}
\end{algorithm}

\textsc{StaticAnalysis} parses the source repository to produce a
repository profile~$\Sigma$ containing file-level definitions,
module structure, call graphs, function/subroutine signatures, I/O
contracts, and file-to-definition mappings. The complete profile is
retained for coverage scoring, but is not sent wholesale to the LLM.
Files are ranked by static-analysis importance, and the prompt receives a
bounded projection of the profile plus excerpts from the highest-ranked
source files. For large repositories, ranked files are grouped into at
most six call-connected module families; each family is processed with
its own projected profile and cross-family interface edges, and the
resulting ADFD fragments are merged. The default prompt budget is
120{,}000 characters. If a projected prompt exceeds its share of this
budget, call edges, procedure and I/O contracts, and source excerpts are
progressively reduced; source excerpts can be removed while a reduced
structural profile is retained. Subsequent coverage loops identify omitted files and
definitions against the complete profile and inject targeted directives.
\textsc{InferADFD} prompts $\lambda$ with this bounded context and any gap
directives ($\mathit{dir}$), returning candidate~$A_i$.
\textsc{CoverageScore} compares it with the profile and returns a scalar
score~$s_i \in [0,1]$ together with a gap set~$\mathit{gaps}_i$ listing
uncovered source files and definitions. The scalar score is file
coverage; uncovered definitions are
retained in $\mathit{gaps}_i$ to generate refinement directives.

\textsc{GapDirectives} converts the gap set into targeted instructions
(e.g., ``\emph{The following source files are not represented: {[}list{]}.
Revise the ADFD to include these.}''), thus
preserving the LLM's freedom to restructure while ensuring omissions are
reported.  In practice, the first pass usually maps 70--85\% of source
files, and the second pass typically exceeds the (configurable) 80\% file-coverage threshold.
Repositories with deeply nested hierarchies, such as M\_strings and xtb, may require further refinement loops.


\subsubsection{Target-Code Generation and Source--Target ADFD Alignment}\label{sec:target-generation}

Procedure~\ref{proc:generate} generates target code chunk by chunk in
topological order and then invokes a separate source--target ADFD alignment step.
The generator LLM~$\lambda_g$ produces code; the evaluation-only judge is
not part of this procedure.
For each chunk~$C_j$,
\textsc{BuildChunkDFD} extracts a \emph{sub-ADFD} from the full
ADFD~$A_S$. It needs the complete ADFD because the chunk's
processes may reference data stores and external entities defined
globally, and flows may connect to elements outside the chunk.
\textsc{UpstreamInterfaces} derives lightweight interface
summaries~$\mathit{ifc}$ from the
accumulated context~$\mathit{ctx}$, which holds the public signatures
of all modules generated by prior chunks.
\textsc{LLMGenerate} receives the sub-ADFD, interface summaries, and
target language, and generates code for the chunk.
Finally, \textsc{Signatures} parses the output to extract public
module signatures appended to~$\mathit{ctx}$ so that subsequent chunks
can reference them correctly.

After each complete pass through all chunks,
\textsc{ExtractTargetADFD} applies Python static analysis to the generated
repository to obtain $A_T$. The function \textsc{AlignDiff} compares
$A_S$ and~$A_T$ to identify structural differences, including missing
and additional processes, flows, and data stores.
This comparison is
deterministic: it normalizes element names, matches source and target
processes and data stores by exact, containment, or token-overlap rules,
and compares flows by their normalized endpoint pairs. It also computes
count-based coverage for processes, flows, and data stores. The resulting
difference set $\mathit{diff}$ supplies structural feedback to
\textsc{LLMGenerate} when each chunk is regenerated. The loop terminates
when no structural differences remain
or when the iteration bound is reached. Appendix~\ref{app:alignment-scoring} gives the
normalization and matching rules in full. Separately, an evaluation-only LLM
judge assesses semantic alignment between $A_S$ and $A_T$ for evaluation
(Appendix~\ref{app:outcome-index}); its score is not used in the
refinement loop.

\begin{algorithm}[t]
\caption{\textsc{GenerateTarget}}\label{proc:generate}
\begin{algorithmic}[1]
\STATE \textbf{Input:} $T$: the target language;\;
        $A_S$: the source ADFD;\;
        $\lambda_g$: a generator LLM;\;
        $c_{\max}$: a bound on processes per chunk;\;
        $m$: a bound on refinement iterations
\STATE \textbf{Output:}  $R_T$: a target repository
\STATE $C \gets \textsc{DependencyChunk}(A_S, c_{\max})$
\STATE $\mathit{diff} \gets \emptyset$
\FOR{$i = 1$ \TO $m$}
\STATE $R_T \gets \emptyset$;\;
       $\mathit{ctx} \gets \emptyset$
    \FOR{\textbf{each} chunk $C_j$ \textbf{in} $C$}
        \STATE $\mathit{sub} \gets \textsc{BuildChunkDFD}(C_j, A_S)$
        \STATE $\mathit{ifc} \gets
            \textsc{UpstreamInterfaces}(\mathit{ctx})$
        \STATE $\mathit{code}_j \gets \textsc{LLMGenerate}(\lambda_g,
            \mathit{sub}, \mathit{ifc}, T, \mathit{diff})$
        \STATE $R_T \gets R_T \cup \mathit{code}_j$
        \STATE $\mathit{ctx} \gets \mathit{ctx} \cup
            \textsc{Signatures}(\mathit{code}_j)$
    \ENDFOR
\STATE $\Sigma_T \gets \textsc{StaticAnalysis}(R_T)$
\STATE $A_T \gets \textsc{ExtractTargetADFD}(R_T, \lambda_g, \Sigma_T, \emptyset)$
\STATE $\mathit{diff} \gets
       \textsc{AlignDiff}(A_S, A_T)$
  \IF {$\mathit{diff} = \emptyset$}
    \STATE \textbf{break}
  \ENDIF
\ENDFOR
\RETURN $R_T$
\end{algorithmic}
\end{algorithm}

The principal difficulty for target-code generation is scale\label{sec:chunking}: real-world ADFDs can
contain dozens of interconnected processes.  Providing the full ADFD to an
LLM as a single prompt usually exceeds the context window and degrades
quality~\cite{liu2024lost}.  We address this with a dependency-aware
chunking procedure \textsc{DependencyChunk}
that partitions the ADFD process graph into bounded,
topologically ordered groups.
Note that this is fundamentally different from the text-based chunking
used in Retrieval-Augmented Generation systems, which splits documents
into fixed-size fragments for retrieval; our chunking operates on a
directed process dependency graph using graph-theoretic algorithms to
preserve referential integrity and generation order.
Given an ADFD and a maximum number of processes per chunk $c_{\max}$,
\textsc{DependencyChunk} is
a multi-stage process, summarized below.

\begin{description}
    \item[Stage 1.]  Dependency Graph Construction.\newline
        Extract the process-to-process dependency graph
        from the ADFD.
    \item[Stage 2.] SCC Detection. Apply Tarjan's algorithm~\cite{tarjan1972depth}
        (iterative variant for large graphs) to find all strongly connected components
        in the graph from Stage 1.
    \item[Stage 3.] Topological Sort. Replace each SCC with a single
        \emph{super-node} that represents all the
        processes it contains, collapsing the mutual dependencies within the SCC.
        The result is now a DAG. We compute a topological ordering of the
        DAG with Kahn's algorithm~\cite{kahn1962topological}, guaranteeing that
        when code is generated for a chunk~$C$, every chunk it depends on has already been
        processed.
    \item[Stage 4.] Greedy Bin-Packing.
    The topologically sorted sequence of SCCs is packed into chunks using a
    greedy strategy, i.e., iterate through the SCCs in topological order.
    When the number of processes in a chunk exceeds the bound
    $c_{\max}$ a new chunk is started.
    Because SCCs are atomic units (never split), a single SCC whose process count
    exceeds $c_{\max}$ simply becomes an oversized chunk on its own, thus
    preserving dependency integrity at the cost of a larger LLM prompt.
\end{description}

The total chunking complexity is $O(n+e)$, where $n=|P|$ and $e$ is the
number of direct process-to-process flows in the ADFD dependency graph.


Once we have the implementations of \textsc{ConstructADFD} and
\textsc{GenerateTarget}, the
porting function is straightforward (see Procedure~\ref{proc:pipeline}).

\begin{algorithm}[t]
\caption{\textsc{Port}}\label{proc:pipeline}
\begin{algorithmic}[1]
\STATE \textbf{Input:} $T$: the target language;\;
        $R_S$: the source repository;\;
        $\lambda_g$: the generator LLM;\;
        $k, \tau, c_{\max}, m$
\STATE \textbf{Output:}  $R_T$: target repository
\STATE $A_S \gets \textsc{ConstructADFD}(R_S, \lambda_g, k, \tau)$
\STATE $R_T \gets \textsc{GenerateTarget}(T, A_S, \lambda_g, c_{\max}, m)$

\RETURN $R_T$
\end{algorithmic}
\end{algorithm}

As part of \textsc{GenerateTarget}, the source ADFD is compared with the
ADFD extracted from the generated target repository, as described above.
Appendix Figure~\ref{fig:minpack-adfd-example} presents the two partial
ADFDs side-by-side for a MINPACK example, providing a visual illustration
of the comparison.

\section{Empirical Evaluation}\label{sec:eval-design}

The principal goal of the experiment is to test the hypothesis:
\begin{quote}
Using an LLM to generate an ADFD as a latent representation of code; and then using the LLM to generate the actual
            code using the ADFD as context results in more effective porting of code repositories
            than direct LLM-based migration of repositories.
\end{quote}
We test this hypothesis on repositories of varying sizes ranging from very simple to very complex, and measure effectiveness along several dimensions that broadly fall into two categories:
\begin{enumerate}
  \item \textbf{Porting soundness.} These measure the
         behavioral agreement between the source and target repositories on identical inputs.
  \item \textbf{Porting completeness.} These measure the extent to which source repositories are
        ported completely into executable and maintainable target repositories.
\end{enumerate}

\subsection{Materials}
\label{sec:mat}

\paragraph{{\tt f2x50}: Benchmark Corpus}
\label{sec:benchmark}

\begin{sloppypar}
To the best of our knowledge, no standard benchmark exists for porting
full Fortran repositories to Python. Existing benchmarks such as CodeXGLUE~\cite{lu2021codexglue}
and HumanEval-X~\cite{zheng2023codegeex} evaluate task-level or
function-level code understanding and generation rather than
repository-scale Fortran porting, and recent repository-level
benchmarks target Java~\cite{ibrahimzada2025alphatrans, may2026freshbrew}.
We fill this gap with {\tt f2x50}, a benchmark designed specifically for
repository-scale
Fortran-to-Python migration. Table~\ref{tab:corpus} summarizes its
three complexity tiers.  It comprises
50 curated open-source Fortran repositories from GitHub, selected for
diversity in domain, coding style, and structural complexity.
\end{sloppypar}

\begin{table}[t]
\caption{{\tt f2x50}: 50 Fortran repository benchmark.}\label{tab:corpus}
\centering
\small
\resizebox{\columnwidth}{!}{%
\begin{tabular}{@{}lcrr>{\raggedright\arraybackslash}p{3.0cm}@{}}
\toprule
\textbf{Tier} & \textbf{$n$} & \textbf{LoC Range} & \textbf{Files} & \textbf{Examples} \\
\midrule
Low     & 20 & 1.5k--41k   & 4--112   & ABAQUS, fftpack, tsunami \\
Medium  & 20 & 4.8k--192k  & 9--405   & minpack, stdlib, xtb \\
Complex & 10 & 29k--1.6M   & 158--3.6k & hdf5, lapack, pymc2 \\
\bottomrule
\end{tabular}
}
\end{table}

\begin{sloppypar}
Repositories span diverse scientific domains: numerical methods
(minpack, quadpack, SISSO), computational fluid dynamics (CaNS, Incompact3d,
Nek5000, openfast), molecular dynamics and quantum chemistry (dftd4,
xtb, crest), astrodynamics (Fortran-Astrodynamics-Toolkit), string and
time processing (M\_strings, M\_time, datetime-fortran), wavelet
analysis (wavelets), testing frameworks (test-drive), weather modeling
(WPS, ccpp-physics), finite element analysis (elmerfem), high-performance
I/O (hdf5), linear algebra (lapack, arpack-ng), Bayesian statistics
(pymc2), package management (fpm), and neural networks (FKB,
neural-fortran, fastGPT).
\end{sloppypar}

Selection criteria: (1)~publicly available on GitHub under an
open-source license; (2)~compilable with a standard Fortran compiler;
(3)~more than 1{,}000~LoC; and (4)~no public Python translation
available, ensuring that no LLM could have been trained on a Python
version of these repositories, making them authentic test cases.
Repositories were stratified by structural
complexity considering inter-module coupling, use of advanced Fortran
features, and codebase scale. Because stratification weighs coupling and
language features as well as size, the LoC ranges of adjacent tiers
overlap (Table~\ref{tab:corpus}).
The benchmark is publicly available at
\url{https://github.com/ShraddhaSurana/iProg/tree/main/f2x50}.

\subsection{Method}
\label{sec:meth}

The steps below were used to test the hypothesis. The ADFD
construction, generation, and alignment steps are the ones specified in
Section~\ref{sec:approach}. Table~\ref{tab:exp-settings} lists the settings used.

\medskip
\begingroup
\small
\setlength{\parindent}{0pt}
\setlength{\parskip}{2pt}

    \textbf{For each} repository class (low, medium, complex):\\
\hspace*{1em}Fix the generator LLM per repository
(Table~\ref{tab:exp-settings}).\\
\hspace*{1em}\textbf{For each} repository in that class:
\begin{enumerate}\setlength{\itemsep}{1pt}\setlength{\leftskip}{1em}
  \item Curate the behavioral probes from the Fortran source.
  \item Construct an ADFD using the construction method of
        Section~\ref{sec:source-adfd}.
  \item Port the code to Python using the ADFD
        (Section~\ref{sec:target-generation}, Procedure~\ref{proc:generate}).
  \item Extract an ADFD from the ported code and compare it with the
        source ADFD using the alignment process of
        Section~\ref{sec:target-generation}. Where the two differ,
        regenerate the code with the differences as feedback and compare
        again. The ADFDs are \emph{aligned} when the comparison reports no
        missing or additional processes, flows, or data stores.
  \item Construct the ported code \emph{without} an ADFD, once for each
        baseline or ablation run on that repository.
  \item Run the fixed probes against every port and against the same
        Fortran oracles, and compute the migration outcome index for
        every port; record both.
\end{enumerate}

\endgroup
\medskip

\noindent
Some additional details are relevant.
Four design choices in these steps control what a difference in outcome
can be attributed to. These are:

\noindent
\emph{(i) Generator assignment.} The generator is assigned by what a model
could handle, not by tier label: the local 30B model ported all 20
low-tier repositories and the 4 medium-tier ones it completed without
excessive stubbing or timeouts, the 480B model the other 16 medium-tier
repositories, and Claude Sonnet~4.5 the 10 complex-tier ones. Whichever
generator a repository was assigned produced both its ADFD-Migrate port
and all of its baseline ports, so a difference in outcome for that
repository is attributable to the method rather than the model.

\noindent
\emph{(ii) Probe provenance.} Probes are curated from the \emph{source} side
only: public, deterministic Fortran routines callable on small
self-contained inputs, with expected values taken from the original
Fortran compiled with \texttt{gfortran} or wrapped with \texttt{f2py}.
They are fixed before any port is produced, so the same probe set is
applied to every port of a repository.

\noindent
\emph{(iii) Baselines and ablations (step~5).} The two baselines are
direct file-by-file translation and repository-context direct
translation; the two ablations are the static-profile ablation (no
ADFD) and the dependency-chunking ablation (no ADFD). All four cover the same 47 repositories: all
low- and medium-tier repositories plus 7 complex-tier repositories
(\texttt{pymc2}, \texttt{ccpp-physics}, \texttt{Nek5000}, \texttt{petsc},
\texttt{hdf5}, \texttt{fpm}, and \texttt{openfast}). The remaining three
largest complex-tier repositories -- \texttt{lapack}, \texttt{cp2k}, and
\texttt{elmerfem} -- could not be tried across all four comparisons due
to cost (e.g., running repository-context direct translation on \texttt{lapack} alone cost
$>\$300$). Together these comparisons separate the contribution
of the declarative representation from that of context and dependency
ordering alone.

\noindent
\emph{(iv) Measures (step~6).} The two measures are defined in
Section~\ref{sec:evaluation}: behavioral agreement is oracle-based and
uses the same probes, oracles, and inputs for every method, whereas the
outcome index is not oracle-based and is computed from each port directly.

\begin{table}[t]
\caption{Experimental settings.}\label{tab:exp-settings}
\centering
\small
\begin{tabular}{@{}ll@{}}
\toprule
\textbf{Setting} & \textbf{Value} \\
\midrule
Generator (low) & Qwen-3-Coder 30B (local) \\
Generator (medium) & Qwen-3-Coder 30B (4), 480B (16) \\
Generator (complex) & Claude Sonnet~4.5 (API) \\
Judge model & Claude Opus~4.6 (evaluation only) \\
Temperature & 0.3 (generation), 0.2 (judge) \\
ADFD iterations $k$ & 3--5 \\
Coverage threshold $\tau$ & 0.80 \\
Chunk size $c_{\max}$ & 3 processes \\
Refinement iterations $m$ & 3--5 \\
\bottomrule
\end{tabular}
\end{table}

\noindent
Appendix~\ref{app:exp-details} gives the remaining details needed to
replicate the experiment: model configurations and timeouts, the prompts
for ADFD inference, target generation, and each baseline, the computation
of the structural and semantic alignment scores, the probe inputs and
tolerances, and the outcome-index formula;
Appendix~\ref{app:reproducibility} reports representative probe outputs
and the computational environment. No developer inspected or
ratified an ADFD in any run reported here; every port was produced
end-to-end without human intervention
(Appendix~\ref{sec:human-in-loop}).

\subsubsection{Evaluation}\label{sec:evaluation}

Step~6 records two measures, corresponding to the two dimensions of
effectiveness stated above.

\paragraph{Porting soundness: behavioral agreement.}
For each probe, the harness executes the Fortran oracle and the
corresponding generated Python on identical inputs and compares the
outputs by numeric tolerance, bit-exact array, or exact scalar
comparison. A probe \emph{passes} if the outputs agree, and \emph{fails}
otherwise. A probe also fails if no corresponding Python behavior can be
located, imported, and invoked: failing to produce runnable target code
for a selected source behavior is a failure of the porting method. Behavioral agreement is therefore reported over all
\emph{planned} probes. Failures of this second kind are additionally
tallied as \emph{missing functionality}, because they diagnose a
different defect (absent or unusable generated code) from a wrong
numerical result, and a method's runnable-only rate is reported alongside
as a diagnostic.

\paragraph{Porting completeness: migration outcome index.}
The index combines ADFD semantic alignment (LLM-judged), executability
(syntax validity and import success), code quality (docstring and
type-hint coverage), and implementation rate (the fraction of generated
functions containing real logic rather than stubs). It measures whether
a method yields a complete, importable, maintainable migration and is
\emph{not} a correctness percentage; correctness claims rest on
behavioral agreement. Appendix~\ref{app:outcome-index} gives the
formula and component definitions.

\section{Results}\label{sec:results}

The hypothesis is supported on both dimensions defined in
Section~\ref{sec:eval-design}: the ADFD-mediated ports reproduce the
source repositories' behavior on nearly four times as many of the planned
probes as direct translation. We report the two dimensions in turn, with
the relevant baselines inside each comparison. Supporting diagnostics follow; their detailed tables are
in Appendix~\ref{app:supporting-results}, and repository-level scores in
Appendix~\ref{app:repository-results}.

\subsection{Porting Soundness: Behavioral Agreement}\label{sec:primary-result}

\noindent
\begin{sloppypar}
{\bf {ADFD-Migrate against the Fortran oracles: }}
The executable Fortran-vs-Python harness planned 382 behavioral probes across all 50 repositories, with at least four
probes per repository.  Every planned probe was runnable against the
ADFD-Migrate ports: 327 passed and 55 failed, with no missing
functionality, no execution errors, and no skipped checks, for a
behavioral agreement rate of 85.6\%; 40 of 50 repositories passed every
probe.  The checks are targeted probes rather than
exhaustive system tests, but they are the direct correctness evidence in
this study: each check compares source and generated behavior
on the same inputs. The tier breakdown is in
Appendix~\ref{app:behavioral-breakdown} and the per-mode counts in
Appendix~\ref{app:probe-curation}; the probes cover numerical, bit-exact,
scalar/string, boolean, and duration-normalized comparisons.
\end{sloppypar}

\noindent
{\bf {ADFD-Migrate against direct translation, same probes: }}
To test whether the intermediate ADFD representation improves porting,
we ran the same 382 curated behavioral probes against the two baselines
and the two ablations, using the same Fortran oracles and
the same input vectors.  The verifier does not require a comparison
method to use the same file layout as ADFD-Migrate: for each probe, it
searches the method's generated \texttt{src/} tree by routine name, so a
routine is still tested if the method relocates it or places it in a
different module. The results are summarized in
Table~\ref{tab:behavioral-by-method}.

Counting every planned probe, ADFD-Migrate agrees with the Fortran
oracles on 85.6\% of the 382 probes, against 21.7\% for both direct
file-by-file translation (83/382) and repository-context direct
translation (83/382), a factor of 3.9, and against 11.0\% for the
static-profile ablation (42/382) and 3.7\% for the dependency-chunking
ablation (14/382). The gap is driven by
runnability: all 382 probes are runnable against ADFD-Migrate, but only
99 and 98 respectively against the two baselines, and only 69 and 30
respectively against the two ablations. The remaining probes
fail as \emph{missing functionality}, which can arise when parts of
a large repository were not generated because of input size constraints,
the relevant routine or wrapper is absent, the generated file has an
import or syntax defect, or the located routine has an incompatible
calling convention.

\begin{table*}[t]
\caption{Cross-method behavioral comparison on the same 382 planned
Fortran-vs-Python probes. \emph{Runnable}: probes for which the generated
Python target could be located, imported, invoked, and compared with the
Fortran oracle. A probe fails either because the generated code returned a
wrong output or because no runnable target existed for it at all
(\emph{missing functionality}: missing generated target, non-generated
repository, import/syntax defect, or incompatible interface). The last two
columns give the same pass counts as a rate over all 382 planned probes,
which is the reported agreement rate, and over only the probes that ran.}
\label{tab:behavioral-by-method}
\centering
\small
\resizebox{\textwidth}{!}{%
\begin{tabular}{@{}lrrrrrrr@{}}
\toprule
\multirow{2}{*}{\textbf{Method}} & \multirow{2}{*}{\textbf{Planned}} &
\multirow{2}{*}{\textbf{Runnable}} & \multirow{2}{*}{\textbf{Pass}} &
\multicolumn{2}{c}{\textbf{Failures}} &
\multicolumn{2}{c}{\textbf{Behavioral agreement}} \\
\cmidrule(lr){5-6} \cmidrule(lr){7-8}
 & & & & \textbf{Wrong output} & \textbf{No runnable target} &
\textbf{Over all 382} & \textbf{Over runnable only} \\
\midrule
ADFD-Migrate            & 382 & 382 & 327 & 55 &   0 & \textbf{85.6\%} & 85.6\% \\
Direct translation      & 382 &  99 &  83 & 16 & 283 & 21.7\% & 83.8\% \\
Repo-context direct     & 382 &  98 &  83 & 15 & 284 & 21.7\% & 84.7\% \\
\midrule
Static profile (no ADFD)      & 382 &  69 &  42 & 27 & 313 & 11.0\% & 60.9\% \\
Dependency chunking (no ADFD) & 382 &  30 &  14 & 16 & 352 &  3.7\% & 46.7\% \\
\bottomrule
\end{tabular}
}
\end{table*}

\begin{sloppypar}
The last two columns of Table~\ref{tab:behavioral-by-method} show where
the advantage comes from. Restricted to the probes a method could
actually run, the two baselines remain comparable to ADFD-Migrate:
85.6\%, 83.8\%, and 84.7\%, with some probes passing only for a baseline
and some only for ADFD-Migrate. So when a baseline does produce runnable
code for a behavior, that code is about as likely to be right. The two
ablations trail on this measure (60.9\% and 46.7\%), indicating
that, unlike the baselines, the routines they do manage to run are also
less often correct. What separates all four comparisons from
ADFD-Migrate is primarily how often they produce runnable code at all,
and this is why the over-all-382 rates diverge so sharply: their high
(or, for the ablations, middling) runnable-only rates are computed over
a much smaller share of the intended behaviors -- roughly a quarter for
the baselines and a fifth or less for the ablations. Because a port
that omits a behavior, or produces code that
cannot be imported or called, has not delivered that behavior to whoever
needs the migration, we treat those probes as failures. The agreement
rate over all 382 planned probes is therefore the one we report, and on
this measure the hypothesis is supported.
\end{sloppypar}

\subsection{Porting Completeness: Migration Outcome Index}
\label{sec:readiness-index-results}

Table~\ref{tab:cross-tier} reports the migration outcome index together
with its principal components and the paired direct-translation
baseline. ADFD-Migrate averages 93.1\% across all 50 repositories and
exceeds direct translation in every tier by 16.9--58.5 percentage points.
The gap grows with repository complexity, which is consistent with the
claim that an explicit approximation of the latent declarative
representation is most useful when file-wise translation loses
repository-wide structure.

\begin{table*}[t]
\caption{Cross-tier migration outcome results. Sem. = ADFD semantic-alignment
score; Syn. = syntax validity; Imp. = import success. Direct comparison
uses 20 low-, 20 medium-, and 7 complex-tier repositories. $\Delta$ is the
ADFD-minus-direct difference in percentage points.}
\label{tab:cross-tier}
\centering
\small
\resizebox{\textwidth}{!}{%
\begin{tabular}{@{}lrrrrrrrr@{}}
\toprule
\textbf{Tier} & \textbf{$n$} & \textbf{Sem.} & \textbf{Syn.} &
\textbf{Imp.} & \textbf{ADFD outcome} & \textbf{Direct $n$} &
\textbf{Direct outcome} & \textbf{$\Delta$} \\
\midrule
Low     & 20 & 95.2\% & 99.9\% & 77.2\% & 93.0\% & 20 & 76.1\% & +16.9 \\
Medium  & 20 & 98.3\% & 98.1\% & 72.5\% & 91.9\% & 20 & 65.3\% & +26.6 \\
Complex & 10 & 99.0\% & 100.0\% & 79.8\% & 95.7\% & 7 & 37.2\% & +58.5 \\
\midrule
\textbf{All} & 50 & 97.2\% & 99.2\% & 75.9\% & \textbf{93.1\%} & & & \\
\bottomrule
\end{tabular}
}
\end{table*}

The stronger repository-context direct baseline improves the plain
direct mean from 0.654 to 0.709, but remains 21.5 percentage points below
ADFD-Migrate and loses on 46 of 47 pairs. The static-profile ablation
and the dependency-chunking ablation also trail ADFD-Migrate on all 47 available pairs
each. These comparisons separate the ADFD contribution
from simply supplying more repository context, static metadata, or
dependency groups. Full baseline and ablation distributions,
representative cases, paired confidence intervals, and tests are
reported in Appendix~\ref{app:extended-baselines}.

\subsection{Supporting Analysis: Mechanism and Practicality}
\label{sec:supporting-summary}

The supporting analyses clarify what the representation does and where
it remains insufficient. Source ADFDs were constructed for all 50
repositories and partitioned into 1--42 dependency-ordered chunks.
Near-perfect syntax validity indicates that bounded generation is
reliable, while import success (72.5--79.8\% by tier) remains the main
integration weakness. A conservative deterministic reconciliation pass
raised mean import success from 79.7\% to 82.3\%
(Appendix~\ref{sec:import-reconciliation}).

Representation-level alignment is useful but not equivalent to
correctness. High outcome-index repositories can still contain
formula-level, formatting, numerical-constant, or numerical-kernel
defects; the four detailed failure modes in
Appendix~\ref{sec:readiness-behavior} motivate adding executable
contracts and source-oracle feedback to the refinement loop. Sensitivity
analysis preserves a positive ADFD--baseline mean gap for 98.9\% of
10{,}000 sampled weightings (Appendix~\ref{sec:sensitivity-results}).

Finally, chunking enabled a local 30B model to port all low-tier
repositories with no metered generation charge. The ten complex
repositories cost about \$117, and the ADFD workflow cost about 54\% of
file-wise direct translation on the seven complex repositories with
recorded paired costs. Appendix~\ref{sec:scaling-results} reports model,
construction, chunking, and error-taxonomy details;
Appendix~\ref{sec:cost-analysis} gives the cost accounting.

\section{Discussion}\label{sec:discussion}

\subsection{Additional Questions and Answers}\label{sec:clarifications}

We now turn to some questions that require additional clarification.
\vspace{1em}

\begin{sloppypar}
\noindent
{\bf {How is data flow recovered, given pointers and aliasing? }}
The ADFD is inferred by an LLM and constrained by a static profile
(Section~\ref{sec:source-adfd}), so no alias analysis is run. Fortran
arguments may not overlap unless declared as pointers or targets, so the
profile's call graph can be trusted as written, and two names for one array
are the same data store regardless. A pointer-heavy language such as C
would first need points-to analysis in its adapter, which needs to be created for each source language.
\end{sloppypar}
\vspace{1em}
\begin{sloppypar}
\noindent
{\bf {Is data flow sufficient, and where does control flow go? }}
Control flow is not discarded; it is captured \emph{inside} the
processes. The diagram
carries only what crosses procedures, which is what chunking and alignment
need. Fidelity to control flow therefore rests on process specification and
contract text, so control-flow-heavy repositories are more challenging: for
example, \texttt{quadpack} (1/12 probes) and \texttt{roots-fortran} (0/7)
recovered the correct processes but not the correct stopping logic
(Appendix~\ref{sec:readiness-behavior}).
\end{sloppypar}

\vspace{1em}
\begin{sloppypar}
\noindent
{\bf {Why not generate the target tests from the source test suite? }}
The probes do this in a stronger form: their expected values come from
executing the original Fortran, and they are fixed before any port exists
(Section~\ref{sec:evaluation}). The existing suites are themselves Fortran
programs, unevenly present across the corpus, so translating one is a
migration task carrying the defect being measured, and a failure would not
separate a faulty port from a faulty test.
\end{sloppypar}

\subsection{Limitations and Threats to Validity}\label{sec:threats}

\begin{sloppypar}
\noindent
{\bf {Construct: }}
The 382 probes are targeted case-level checks, not exhaustive system
tests, so behavioral agreement bounds correctness from below rather than
certifying it. Exhaustive equivalence testing is not currently feasible
across this corpus: many repositories need large scientific inputs,
MPI/HDF5 or compiler-specific runtimes, and project-specific output
oracles that do not exist uniformly. The outcome index is a separate
matter: it encodes a preference for complete, importable, maintainable
draft migrations, and sensitivity analysis supports its comparative use
without making it a correctness percentage. The two ablations
show the ADFD contributes beyond static metadata and dependency-only
chunking, but a component-by-component ablation remains future work.
Human expert verification of the ported code has not been possible: the
benchmark spans scientific domains (e.g., quantum chemistry, finite
element analysis, weather modeling) that require specialized expertise
the authors do not have and this expertise has been difficult to find,
 so probe-based and LLM-judged evaluation are
the only correctness signals reported here.
\end{sloppypar}

\noindent
{\bf {Internal: }}
Temperature~0.3 makes generation non-deterministic; we mitigate this by
evaluating 50 diverse repositories rather than repeated runs of a few.
The Claude Opus~4.6 judge is independent of the Qwen generators but
shares a model family with the complex-tier Sonnet~4.5 generator, so its
scores may carry family bias; no correctness claim depends on it. Prompt
effort favors the baselines and ablations, whose prompts were hand-refined to their
best obtainable results while ADFD-Migrate ran with fixed prompts and no
per-repository tuning (Section~\ref{sec:meth}).

\begin{sloppypar}
\noindent
{\bf {External: }}
Only Fortran-to-Python porting was evaluated, and the approach suits
numerically oriented software where data-flow structure dominates
control-flow complexity; heavy I/O, event-driven, or GUI logic may
benefit less. The ADFD and chunking algorithms are language-agnostic,
but construction needs language-specific analyzers, generation needs
target-language instructions, and import reconciliation is Python-specific,
so a pair such as COBOL-to-Java requires new adapters. The three model
configurations (Qwen-3-Coder 30B, Qwen-3-Coder 480B, and Claude
Sonnet~4.5) give some evidence of model-agnosticism.
\end{sloppypar}

\section{Conclusion}\label{sec:conclusion}

ADFD-Migrate starts from the premise that a program implements an
unobserved declarative description of its computation. It makes an
inferred ADFD approximation of this latent declarative representation
explicit and inspectable, then uses dependency-aware chunking to generate
a target repository in bounded, ordered units. On the
50-repository {\tt f2x50} benchmark, it passes 327 of 382 source-oracle
behavioral probes and exposes 3.9$\times$ as many planned behaviors as
runnable targets as either direct baseline. Its 17--59 percentage-point
outcome-index advantage is consistent across repository tiers, although
behavioral failures show that process-level alignment cannot guarantee
formula-level fidelity.

Three insights emerge. First, an ADFD can make a useful approximation of
the latent declarative representation explicit, inspectable, and
language-agnostic. Second, dependency-aware chunking enables smaller
open-source LLMs to generate code. Third, structuring the problem via semantic decomposition offers
an alternative to relying only on model scale. Future work will
integrate failed source-oracle probes into chunk refinement, strengthen
domain-specific numerical contracts, evaluate smaller models and additional
language pairs, and develop richer domain-aware repair beyond import
reconciliation.

\section*{Data Availability Statement}
\begin{sloppypar}
The {\tt f2x50} benchmark corpus catalog and reproduction scripts are available at
\url{https://github.com/ShraddhaSurana/iProg/tree/main/f2x50}. The inferred ADFDs, generated Python, metrics, manifests, and analysis outputs
are available at \url{https://doi.org/10.5281/zenodo.22054047}.
\end{sloppypar}


\section*{CRediT authorship contribution statement}
\textbf{Shraddha Surana:} Conceptualization, Methodology, Software,
Investigation, Formal analysis, Data curation, Writing - original draft,
Writing - review and editing.
\textbf{Ashwin Srinivasan:} Conceptualization, Supervision,
Writing - review and editing.
\textbf{Michael Bain:} Writing - review and editing.


\section*{Declaration of competing interests}
The authors declare that they have no known competing financial interests
or personal relationships that could have appeared to influence the work
reported in this paper.


\section*{Funding}
This research did not receive any specific grant from funding agencies
in the public, commercial, or not-for-profit sectors.


\appendix

\renewcommand{\thefigure}{\thesection.\arabic{figure}}
\renewcommand{\thetable}{\thesection.\arabic{table}}
\renewcommand{\theHfigure}{\thesection.\arabic{figure}}
\renewcommand{\theHtable}{\thesection.\arabic{table}}

\section{Full ADFD-Migrate Architecture}
\label{app:full-architecture}
\setcounter{figure}{0}

\begin{figure*}[!t]
\centering
\includegraphics[width=\textwidth]{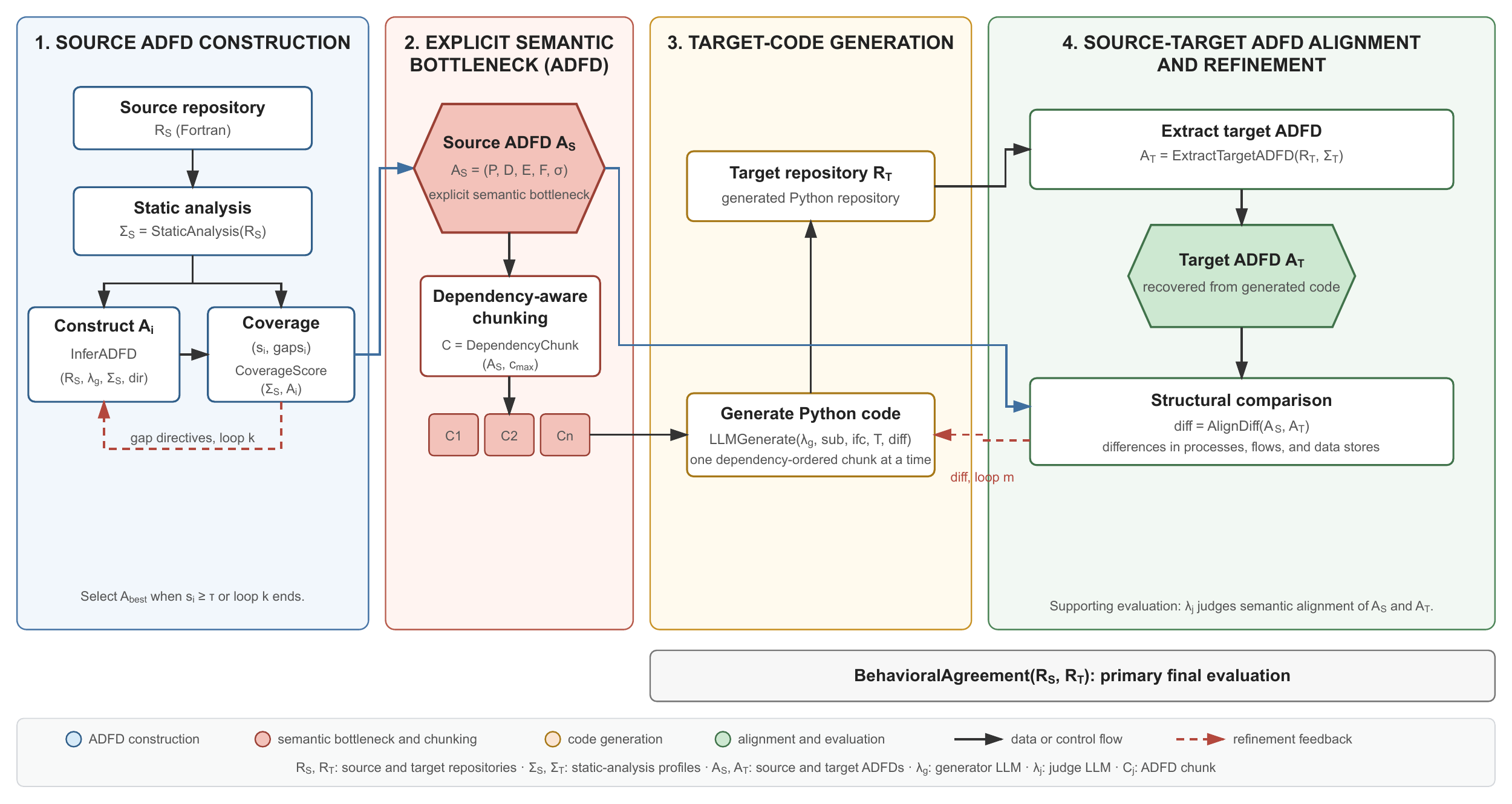}
\caption{Full ADFD-Migrate system architecture. Static-analysis coverage
refines the inferred source ADFD; dependency-aware chunks guide generation; source--target ADFD differences
guide regeneration. The LLM judge supplies supporting
representation-level evidence. Behavioral agreement is the primary evaluation.}
\label{fig:full-architecture}
\end{figure*}

Figure~\ref{fig:full-architecture} maps the porting workflow of
Figure~\ref{fig:pipeline} to the implemented system, expanding its three
steps (convert source code to an ADFD, inspect or revise the ADFD,
generate target code from it) into the automatic components used in the
experiments.

\paragraph{Recovering the source representation.}
The left-hand stage implements $A_S=\mathit{Construct}_S(R_S)$. A
deterministic static-analysis agent parses $R_S$ to produce the
repository profile $\Sigma_S$: files, definitions, module relationships,
call edges, procedure signatures, and input/output information. The
profile is used in two ways. A bounded projection of it, together with
ranked source excerpts, conditions the LLM that infers a candidate source
ADFD. The complete profile is retained by the coverage agent, which
checks which source files and definitions that candidate represents; if
coverage is below threshold $\tau$, the uncovered elements become focused
gap directives for the next inference pass. The
best candidate is retained, and the loop stops at the threshold or after
the configured iteration bound. Static analysis therefore
constrains and checks the explicit ADFD approximation produced by the LLM.

\paragraph{Using the ADFD as a semantic bottleneck.}
The middle stage begins from
$A_S=(P,D,E,F,\sigma)$, rather than translating the source files
independently. A deterministic dependency agent projects the
process-to-process flows, collapses strongly connected components, and
topologically orders the resulting graph. It then packs that order into
chunks $C_1,\ldots,C_n$, subject to the process bound $c_{\max}$ while
keeping each strongly connected component intact. For each chunk, the
generator receives its sub-ADFD, the contracts of referenced data stores
and external entities, and interface summaries from already generated
chunks. This is the operational role of the latent-representation
claim: the inferred ADFD provides a language-independent,
repository-level account of the computation that mediates between the
observed Fortran repository and the generated Python repository.

\paragraph{Checking and refining the target.}
After each generation pass, deterministic Python analysis produces profile $\Sigma_T$
and target ADFD $A_T$. Differences between $A_S$ and $A_T$ guide the next
bounded regeneration pass. This loop tests whether the generated
repository realizes the computational structure made explicit in $A_S$. The evaluation-only LLM judge supplies a
post-generation semantic score and is not part of this loop.

\paragraph{Evaluation boundary.}
The bottom path distinguishes system checks from the final experimental
criterion. Coverage and source--target ADFD alignment measure whether the
representation is complete enough to guide generation and whether its
structure is preserved; they cannot establish behavioral correctness. The
primary test therefore executes matched source-oracle probes against
$R_S$ and $R_T$. The ADFD is the proposed mechanism for improving
porting, structural and semantic alignment diagnose that mechanism, and
observable Fortran--Python agreement evaluates the ported program.

\section{Worked ADFD Example: MINPACK}
\setcounter{figure}{0}

\begin{figure*}[!t]
\centering
\textbf{(a) Five-process source-ADFD cluster inferred from Fortran MINPACK}\par\smallskip
\includegraphics[width=0.94\textwidth]{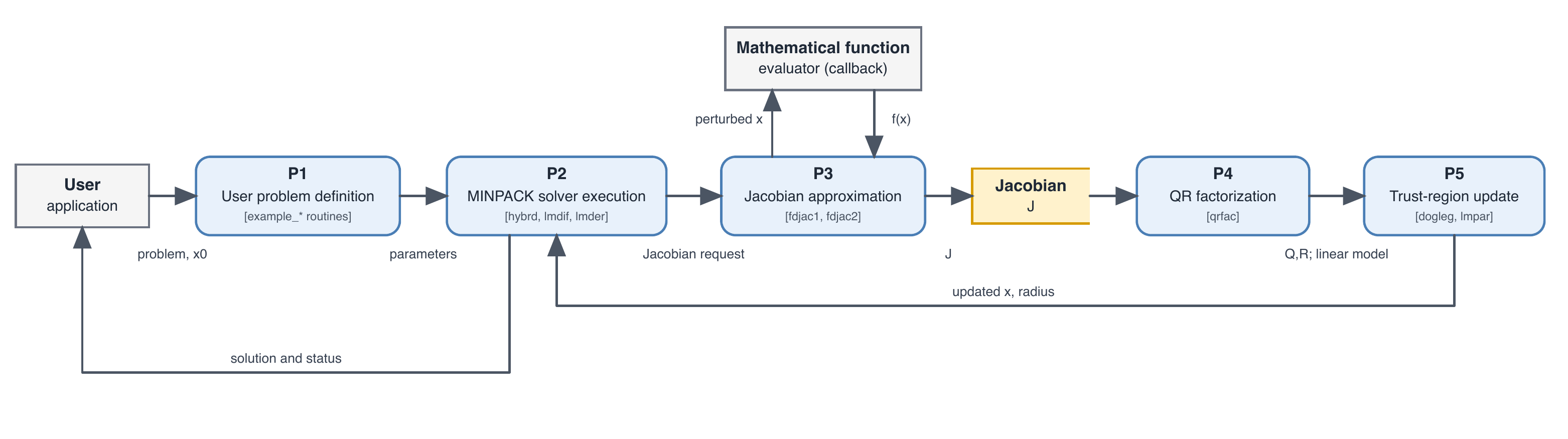}\par\medskip
\textbf{(b) Corresponding excerpt from the generated-Python ADFD}\par\smallskip
\includegraphics[width=0.82\textwidth]{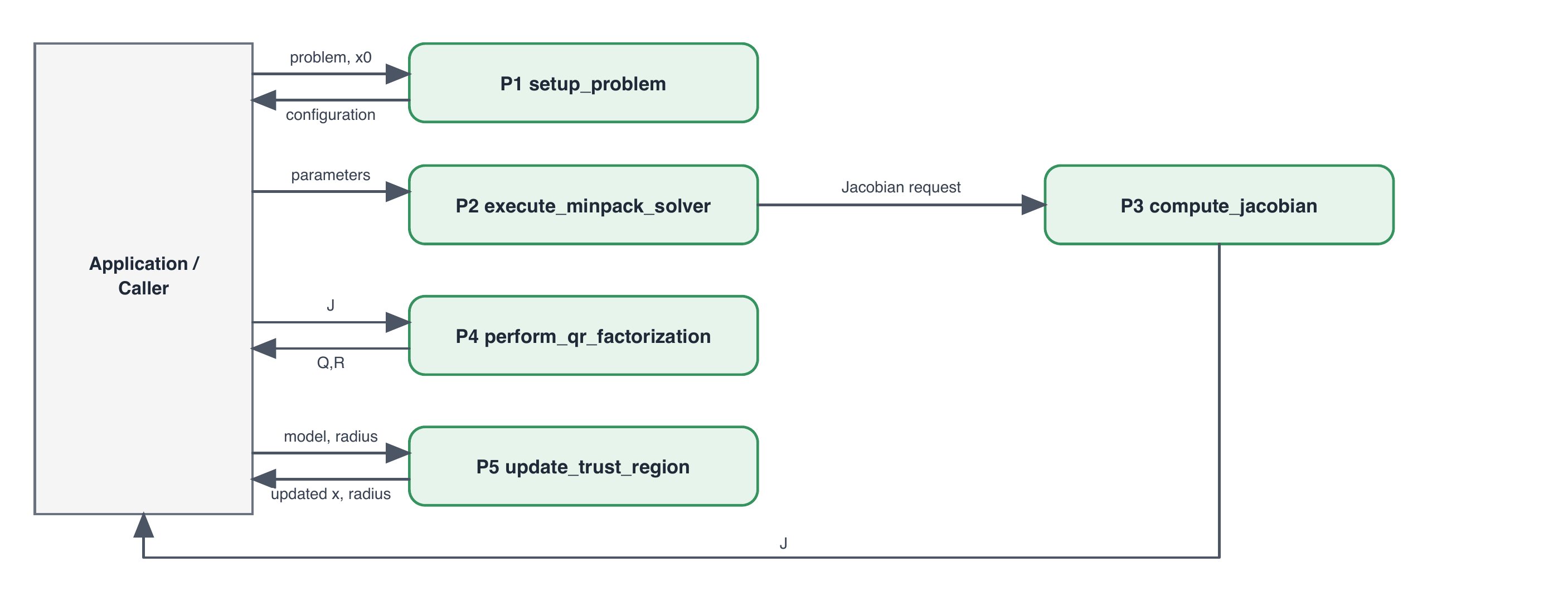}
\caption{Worked MINPACK ADFD excerpt using Gane--Sarson notation.
Rounded rectangles are processes, the open-ended rectangle is a data store,
square-cornered rectangles are external entities, and labeled arrows are flows.
P1--P5 identify corresponding computational roles in
the source and generated-Python ADFDs.}
\label{fig:minpack-adfd-example}
\end{figure*}

Figure~\ref{fig:minpack-adfd-example} shows one five-process numerical
cluster from \texttt{minpack}, rather than its complete ADFD. The source
artifact contains 16 processes, 72 flows, 10 data stores, and four
external entities and is generated in three chunks; the refined target ADFD
contains 32 extracted processes and 48 flows. Panel~(a) maps representative
MINPACK routines to ADFD processes, with the Jacobian as a data store and
the user application and callback as external entities. Panel~(b) uses
process names extracted from the generated Python, enabling direct
comparison of computational roles and flows.

\section{Experimental Details}
\label{app:exp-details}
\setcounter{table}{0}

This appendix provides the detailed experimental configuration,
baseline and ablation descriptions, probe curation methodology, and outcome index
definition referenced in Section~\ref{sec:meth}.

\subsection{Model Configurations}

\paragraph{Generation models.}
Three generator configurations were used. Qwen-3-Coder 30B, hosted locally via Ollama, ported all 20
low-tier repositories and 4 of the 20 medium-tier ones (wavelets, SISSO,
quadpack, MPM3D-F90); Qwen-3-Coder 480B via cloud-hosted Ollama ported the
other 16 medium-tier repositories, on which the 30B model produced
excessive stubs or timed out; Anthropic Claude Sonnet~4.5 via API ported
all 10 complex-tier repositories. Each repository's generator was also used
for all of its baseline and ablation ports.

All generation models used temperature~0.3, JSON response format, and
1200\,s timeout. The non-zero temperature allowed bounded refinement
retries to explore alternatives after a repeated structural defect.
We did not tune temperature; 0.3 is a fixed setting, not an empirical optimum.

\paragraph{Judge model.}
Claude Opus~4.6 (extended thinking; temperature~0.2; 600\,s timeout) was
used only for semantic evaluation. It is independent of the Qwen
generators, but shares the Anthropic family with the complex-tier
generator. Target-ADFD label enrichment also uses the generator, so the
score may retain family and construction bias~\cite{panickssery2024selfpreference}.
This score is supporting evidence only.

\subsection{Baseline and Ablation Descriptions}

\paragraph{Direct baseline.}
Translates each Fortran file to Python independently
without an ADFD-mediated intermediate representation.

\paragraph{Repository-context baseline.}
Additionally supplies neighboring modules and dependency summaries
while retaining independent file generation.

\paragraph{Static-profile ablation (no ADFD).}
A pure static-analysis baseline: it supplies the same repository
profile used to construct the source ADFD (module structure, call
graphs, function/subroutine signatures, and file-to-definition
mappings) but omits the ADFD itself, isolating the contribution of the
declarative representation on top of static analysis alone.

\paragraph{Dependency-chunking ablation (no ADFD).}
Chunks the source call graph using strongly connected component
detection and topological ordering, the same dependency-ordering
mechanism ADFD-Migrate uses, but applied directly to the call graph
instead of an ADFD process graph. This isolates the contribution of the
declarative representation on top of dependency-aware chunking alone.

\subsection{Alignment Scoring}
\label{app:alignment-scoring}

The source ADFD~$A_S$ and the ADFD~$A_T$ extracted from the generated code
are compared twice: a deterministic structural comparison that drives the
refinement loop, and an LLM-judged semantic comparison used only for
evaluation.

\paragraph{Structural alignment (\textsc{AlignDiff}).}
Names are lower-cased and non-alphanumeric characters become
underscores. Normalized names match by equality, substring containment, or
Jaccard overlap $|a \cap b|/|a \cup b| \ge 0.5$ between their
underscore-delimited token sets. Thus, \texttt{solve\_linear\_system} and
\texttt{linear\_system\_solver} match at $2/4$. Flows are normalized
(source, target) pairs. Unmatched processes, flows, and data stores on either
side (i.e. missing or extra) form $\mathit{diff}$, which Procedure~\ref{proc:generate} passes to
\textsc{LLMGenerate}, and
their empty union is the alignment criterion of Section~\ref{sec:meth}.

The comparison also reports rename-insensitive count-coverage and
topology-similarity summaries, combined into a composite structural
alignment score. These are diagnostic only: the refinement loop reacts to
the missing and extra sets, never to the composite score, so a port is
never regenerated merely for renaming an element.

\paragraph{Semantic alignment ($S_{\text{sem}}$).}
The judge receives both ADFDs as text: each process with its name,
specification, and declared inputs and
outputs; the flow list with endpoint names and labels; and the data-store
and external-entity names. It is instructed that names will differ across
languages and must not be penalized on that basis, that one source process
may legitimately map to several target processes or the reverse, and that
each aligned pair is scored on $[0,1]$ with $1.0$ identical functionality,
$0.5$ partial overlap, and $0.0$ unrelated. It returns the aligned pairs
with per-pair scores, the unmatched processes on each side, and one overall
score. $S_{\text{sem}}$ is that overall score clamped to $[0,1]$.

\subsection{Probe Curation}
\label{app:probe-curation}

Probes were selected from the \emph{source} side: routines were chosen
from the Fortran repository for being public, deterministic, and
executable with small self-contained inputs. Each probe's expected
behavior is defined entirely by the Fortran oracle, not by any generated
output. Probe selection was therefore independent of which migration
method produced the Python under test.

Where a generated repository exposes a source behavior under a
different name or module (e.g., \texttt{string\_to\_value}
realized as \texttt{atoi\_int32}), the harness resolves the target by
name-based search over the generated \texttt{src/} tree and, where
needed, repository-specific driver adapters; this resolution step is
applied identically to baseline and ADFD-Migrate outputs.

Every probe is curated against an executable oracle: either the original
Fortran routine compiled with \texttt{gfortran} or a reference \texttt{f2py}
stub (a Python-callable wrapper compiled directly from the original Fortran
via \texttt{f2py}~\cite{peterson2009f2py}), run on the same inputs as the
generated Python.

\paragraph{Comparison modes and tolerances.}
Each probe records its inputs, one of five comparison modes, and numeric
tolerances (absolute and relative) where applicable. Of 382 probes, 205 use numeric tolerance, 93
array equality, 57 scalar/string equality, 23 Boolean equality, and four
equality after unit normalization. Numeric agreement uses
\texttt{numpy.allclose}:
$|x_{\text{py}} - x_{\text{f}}| \le \mathit{atol} +
\mathit{rtol}\,|x_{\text{f}}|$ elementwise with matching shapes for arrays. Per-probe
tolerances are $10^{-14}$ for closed-form functions, $10^{-12}$ for BLAS
kernels, $10^{-10}$ or $10^{-8}$ for composite routines and table lookups,
and $10^{-6}$ for iterative solvers; exact modes use zero tolerance. Inputs are small, self-contained, and
written into the probe definition, so oracle and port receive identical
values.

\subsection{Prompts}
\label{app:prompts}

\paragraph{ADFD inference.}
\textsc{InferADFD} runs two LLM passes per module family: pass~1 produces
the structure, validated without requiring per-process detail, and pass~2
fills in specifications, conditions, and typed I/O, validated against the
expected source-file list. The per-family fragments then receive disjoint ID
prefixes and are merged. The system prompt casts the model as \emph{``a
principal software architect building a reconstruction-grade data flow
diagram''} (pass~2 adds \emph{``and QA engineer refining a DFD for
cross-language repository reconstruction''}) and requires strict JSON with
no markdown.

Each user prompt is a JSON payload: the task statement, the repository name
and source language, the constraints, the bounded profile projection, and,
from the second construction iteration onward, the gap directives and a
summary of the previous best ADFD. The constraints fix the schema (each
process carries an id, name, type, specification, pre- and post-conditions,
typed inputs and outputs, and the list of source files it covers, which is
what makes coverage scoring possible; each flow carries an id, source,
target, and label) and require stable ID prefixes, flows referencing only
existing IDs, at least one flow per process, explicit flow labels, and
specifications of at least 40 characters.

\paragraph{Target generation.}
System prompt: \emph{``You are a principal Python engineer specializing
in porting numerical and scientific Fortran libraries to Python. Given
an Annotated Data Flow Diagram (ADFD), generate a complete,
well-structured Python implementation.''} Requirements instruct the
model to use NumPy/SciPy where appropriate, preserve numerical precision
(\texttt{np.float64} by default), include type hints and docstrings,
respect the ADFD pre/post-conditions, and return only JSON of the form
\texttt{\{"files": [\{"path":\ldots, "content":\ldots\}]\}}. For chunked
generation the prompt additionally states the chunk index, the total
chunk count, and that the chunk may import from earlier chunks listed in
\texttt{upstream\_modules} but not the reverse.

\paragraph{Direct baseline.}
System prompt: \emph{``You are an expert Fortran-to-Python translator.
Given a Fortran source file, produce the equivalent Python code that
faithfully reproduces the same functionality.''} Rules instruct the
model to preserve computational logic exactly, use idiomatic Python,
include type hints and docstrings, use \texttt{snake\_case}, and return
only valid Python. The user prompt supplies the file name and its full
source.

\paragraph{Repository-context baseline.}
One prompt per source file: \emph{``Translate one Fortran source file to
Python using repository context\ldots Do not use, infer any declarative intermediate
representation. Use the repository context for module names,
interfaces, dependencies, shared constants, and naming.''} The prompt gives
the repository name, source path, expected output path, a compact
repository context as JSON, and the full Fortran source, and requires one
primary Python file at the expected path with complete implementations,
imports, and preserved public routine names.

\paragraph{Static-profile ablation (no ADFD).}
One prompt per repository: \emph{``You are translating an open-source
Fortran repository to Python. Use the repository-level static-analysis
profile below. Generate a Python
repository that preserves the observable behavior, public routines, module
structure, and dependencies as much as possible.''} The compacted profile
follows as JSON, file summaries ordered by static importance and fitted to
the same character budget as the ADFD prompts.

\paragraph{Dependency-chunking ablation (no ADFD).}
One prompt per chunk: \emph{``Translate this call-graph/SCC chunk of a
Fortran repository to Python. Use only graph chunking, static context, and upstream
interface summaries.''} The prompt lists the chunk's nodes and the upstream
interface summaries as JSON and asks for complete Python files preserving
callable interfaces. Chunks come from the same SCC detection and
topological ordering as ADFD-Migrate (Stages~1--3 of
\textsc{DependencyChunk}), applied to the source call graph rather than to
an ADFD process graph.

\paragraph{Baseline and ablation generation harness.}
All four share one assembly system prompt (\emph{``You are a
principal Python engineer specializing in repository-scale
Fortran-to-Python migration for scientific software''}) that fixes the same
JSON file-list output schema, path conventions, and engineering
requirements as ADFD-Migrate. They therefore differ only in the context
they receive. The complete prompt texts are in the replication package.

\subsection{Why Not Reference-Based Metrics}
\label{app:no-reference-metrics}

Text-similarity metrics such as BLEU~\cite{papineni2002bleu} and
CodeBLEU~\cite{ren2020codebleu} are inapplicable because no reference
Python translation of these repositories exists, which is a selection
criterion of {\tt f2x50} (Section~\ref{sec:benchmark}). Even given
references, cross-language ports legitimately differ in syntax and idiom,
and prior work reports poor correlation between these metrics and
functional correctness~\cite{macedo2025intertrans,
ibrahimzada2025alphatrans}. The same property rules out training-data
contamination: there is no Python version for a model to have memorized.

\subsection{Migration Outcome Index}
\label{app:outcome-index}

For comparison with baselines, we use a weighted migration outcome index:
\[
  \text{Outcome} = \bigl(0.6 \cdot S_{\text{sem}}
                   + 0.3 \cdot S_{\text{exec}}
                   + 0.1 \cdot S_{\text{qua}}\bigr)
                   \;\times\; r_{\mathrm{impl}}
\]

\paragraph{Component definitions.}
\begin{itemize}
\item $S_{\text{sem}}$: LLM-judged ADFD semantic-alignment score
      (Appendix~\ref{app:alignment-scoring}).
\item $S_{\text{exec}}$: Executability score, calculated as
      $(0.6 \cdot F_{\text{syn}} + 0.4 \cdot F_{\text{imp}})$, where
      $F_{\text{syn}}$ is the fraction of generated \texttt{.py} files
      parseable by \texttt{ast.parse} and $F_{\text{imp}}$ is the fraction
      that execute cleanly when loaded via \texttt{importlib} after the
      generated \texttt{requirements.txt} has been installed.
\item $S_{\text{qua}}$: Code quality, calculated as
      $\tfrac{(C_{\text{doc}} + C_{\text{type}})}{2}$ over all generated
      function definitions, where $C_{\text{doc}}$ is the fraction whose
      body opens with a string literal and $C_{\text{type}}$ the fraction
      carrying a return annotation or at least one annotated parameter.
\item $r_{\mathrm{impl}}$: Implementation rate, the proportion of generated
      functions whose bodies contain real logic. A function counts as a
      stub when, after an optional docstring, its body is exactly one of:
      \texttt{pass}, \texttt{...}, a \texttt{raise} statement, a bare
      \texttt{return}, or \texttt{return None}.
\end{itemize}

For fully implemented ports $r_{\mathrm{impl}}\approx 1$ and the multiplier
has no effect; for scaffold-only conversions it penalizes the score
proportionally.

\subsection{Generated Tests}

Generated pytest suites execute the generated Python, but their expected
outcomes are generated from the target-side code or from generic testing
patterns rather than from an independent Fortran oracle. They expose
syntax, import, packaging, and runtime failures and provide regression
scaffolding.

\subsection{Structural Comparison}

In addition to the ADFD alignment in
Appendix~\ref{app:alignment-scoring}, a structural score compares
AST-level attributes (function count, class count, identifier overlap, and
call-graph edges) between the generated Python and the original Fortran.
It is \emph{not} included in the outcome index and does not drive
refinement, because cross-language porting legitimately changes
identifiers and module boundaries (for example, replacing Fortran COMMON
blocks with Python classes). It is recorded as a diagnostic only.

\section{Supporting Results and Diagnostics}
\label{app:supporting-results}
\setcounter{table}{0}

\subsection{Behavioral Agreement by Tier}
\label{app:behavioral-breakdown}

Table~\ref{tab:behavioral-validation} decomposes the 382 source-oracle
probes by repository tier.

\begin{nonfloattable}
\caption{Executable Fortran-vs-Python behavioral validation by tier.}
\label{tab:behavioral-validation}
\centering
\small
\begin{tabular}{@{}lrrrrr@{}}
\toprule
\textbf{Tier} & \textbf{Repos} & \textbf{Checks} & \textbf{Pass} &
\textbf{Fail} & \textbf{Rate} \\
\midrule
Low & 20 & 142 & 116 & 26 & 81.7\% \\
Medium & 20 & 127 & 110 & 17 & 86.6\% \\
Complex & 10 & 113 & 101 & 12 & 89.4\% \\
\midrule
\textbf{All} & 50 & 382 & 327 & 55 & \textbf{85.6\%} \\
\bottomrule
\end{tabular}
\end{nonfloattable}

\subsection{Extended Baseline and Ablation Comparisons}
\label{app:extended-baselines}

Table~\ref{tab:alternative-baselines} gives the full outcome-index
distributions for ADFD-Migrate and the four non-ADFD alternatives
compared in Section~\ref{sec:readiness-index-results}.

\begin{nonfloattable}
\caption{Migration outcome comparison with non-ADFD alternatives.
All four comparisons share a common 47-repository subset.}
\label{tab:alternative-baselines}
\centering
\small
\resizebox{\columnwidth}{!}{%
\begin{tabular}{@{}lrrrr@{}}
\toprule
\textbf{Method} & \textbf{$n$} & \textbf{Mean} & \textbf{Median} & \textbf{Range} \\
\midrule
ADFD-Migrate              & 47 & 0.924 & 0.939 & 0.771--0.989 \\
Direct translation        & 47 & 0.654 & 0.764 & 0.000--0.980 \\
Repository-context direct & 47 & 0.709 & 0.729 & 0.397--0.924 \\
\midrule
Static profile (no ADFD)      & 47 & 0.498 & 0.508 & 0.000--0.741 \\
Dependency chunking (no ADFD) & 47 & 0.535 & 0.517 & 0.153--0.881 \\
\bottomrule
\end{tabular}}
\end{nonfloattable}

Table~\ref{tab:baseline-cases} shows representative repository-level
differences.

\begin{nonfloattable}
\caption{Representative repository-level migration outcome comparisons.}
\label{tab:baseline-cases}
\centering
\small
\begin{tabular}{@{}lrrr@{}}
\toprule
\textbf{Repository} & \textbf{Direct} & \textbf{ADFD} & \textbf{$\Delta$ (pp)} \\
\midrule
CaNS          & 42.1\% & 94.2\% & +52.1 \\
fftpack       & 61.4\% & 92.0\% & +30.6 \\
toml-f        & 60.0\% & 85.5\% & +25.5 \\
F-A-Toolkit   & 83.9\% & 95.7\% & +11.8 \\
arpack-ng     & 38.5\% & 96.3\% & +57.8 \\
dftd4         & 44.2\% & 95.3\% & +51.1 \\
\bottomrule
\end{tabular}
\end{nonfloattable}

Across all available pairs, ADFD-Migrate improves over direct
translation by 0.270 mean outcome-index points (95\% bootstrap
confidence interval [0.206, 0.341]), over the static-profile ablation
by 0.426 [0.374, 0.476], over the dependency-chunking ablation by 0.389 [0.343, 0.435],
and over repository-context direct by 0.215 [0.181, 0.250]. Wilcoxon
signed-rank tests are consistent with positive improvements
($p < 0.001$).

\subsection{Generated Tests and Import Reconciliation}
\label{sec:import-reconciliation}

Generated tests produced 1,232 executions: 1,205 passed, one failed, and
26 were skipped because target-side defects prevented execution. Among
the 1,206 non-skipped tests, the low, medium, and complex tiers passed
452/453, 509/509, and 244/244 respectively (99.9\% overall). These
tests provide runtime sanity checks but are not independent evidence of
Fortran equivalence.

Of 972 generated modules, 770 imported successfully and 202 failed.
Missing dependencies (71, 35\%) and
\texttt{Import\allowbreak Error} (47, 23\%)
dominate, followed by circular-import symptoms (22, 11\%),
\texttt{ModuleNotFound\allowbreak Error} (21, 10\%), missing-symbol
\texttt{Name\allowbreak Error} (21, 10\%),
\texttt{Attribute\allowbreak Error} (13, 6\%), and
\texttt{Syntax\allowbreak Error} (7, 3\%). A conservative AST-based reconciliation
pass modified 37 files in copied outputs and improved 12 repositories,
raising mean module-level import success from 79.7\% to 82.3\%. It repairs verifiable
module paths, re-exports, package initializers, and standard-library
imports, but does not synthesize logic or repair behavioral defects.

\subsection{Outcome Index Versus Behavioral Agreement}
\label{sec:readiness-behavior}

Outcome index and behavioral agreement are not monotonic. The mean
outcome index is 93.5\% for the 40 repositories with no failed probes,
90.7\% for the two with 1--2 failures, 92.1\% for the four with 4--5
failures, and 91.7\% for the four with 7--11 failures. The
worst-agreement cases show what representation-level measures miss:
\begin{itemize}
  \item \texttt{roots-fortran} (97.0\% outcome) has a
        formula-level error in bisection termination that recurs across
        its Brent, Ridders, and Illinois solvers;
  \item \texttt{Incompact3d} (91.6\%) uses incorrect Pad\'{e}
        stencil coefficients;
  \item \texttt{quadpack} (81.1\%) disagrees in the \texttt{dqk15}
        and \texttt{dqk21} numerical kernels; and
  \item \texttt{ccpp-physics} (96.6\%, 7/11) differs in Fortran
        diagnostic formatting and side effects.
\end{itemize}
These defects motivate richer ADFD contracts with boundary behavior,
formatting conventions, numerical examples, and source-oracle probes
inside the refinement loop. They are also the kind of localized,
contract-level issue that the human ratification checkpoint in
Figure~\ref{fig:pipeline} is designed to let a domain expert catch
before generation; that interaction is not evaluated in this paper.

\subsection{Sensitivity of the Migration Outcome Index}
\label{sec:sensitivity-results}

Table~\ref{tab:sensitivity} summarizes alternative aggregation rules.
In 10{,}000 random Dirichlet weightings, 98.9\% preserve a positive mean
ADFD--baseline gap; the mean sampled gap is 0.143 and the median is
0.146.

\begin{nonfloattable}
\caption{Sensitivity of the migration outcome comparison.}
\label{tab:sensitivity}
\centering
\small
\resizebox{\columnwidth}{!}{%
\begin{tabular}{@{}lrrrr@{}}
\toprule
\textbf{Configuration} & \textbf{ADFD} & \textbf{Direct} & \textbf{Gap} & \textbf{ADFD $>$ Direct} \\
\midrule
Paper weights            & 0.926 & 0.660 & 0.263 & 95.7\% \\
Drop implementation gate & 0.941 & 0.761 & 0.179 & 91.5\% \\
Geometric mean           & 0.923 & 0.633 & 0.288 & 95.7\% \\
Weakest-link minimum     & 0.826 & 0.534 & 0.285 & 80.9\% \\
Add structural 15\%      & 0.897 & 0.681 & 0.213 & 89.4\% \\
\bottomrule
\end{tabular}}
\end{nonfloattable}

\subsection{Model Scale, ADFD Construction, and Generation}
\label{sec:scaling-results}

The local 30B model ported all 20 low-tier repositories with a 93.0\%
mean outcome index, compared with 76.1\% for the same model under direct
translation. Within the medium tier, the four repositories completed by
30B average 83.6\%, and the 16 completed by 480B average 94.0\%; this is
descriptive because model selection depended on successful completion.

Source ADFDs contain 1--53 processes in the low tier (mean 19.1), 9--132
in the medium tier (mean 27.4), and 22--85 in the complex tier (mean
41.1). With $c_{\max}=3$, these become 1--42 chunks. Large SCCs are never
split, preserving dependency correctness at the cost of occasional
oversized chunks. Table~\ref{tab:error-taxonomy} records the generated-code
error categories used in diagnosis.

\begin{nonfloattable}
\caption{Generated-code error taxonomy.}
\label{tab:error-taxonomy}
\centering
\scriptsize
\begin{tabular}{@{}clll@{}}
\toprule
\textbf{ID} & \textbf{Category} & \textbf{Severity} & \textbf{Detection} \\
\midrule
E1 & API hallucination        & Major    & AST/import check \\
E2 & Type mismatch            & Major    & Static pattern \\
E3 & Control-flow divergence  & Major    & LLM judge \\
E4 & Interface mismatch       & Minor    & ADFD contract diff \\
E5 & Numerical precision      & Minor    & Behavioral probe \\
E6 & Syntax                   & Critical & \texttt{ast.parse} \\
E7 & Import failure           & Critical & Runtime import test \\
\bottomrule
\end{tabular}
\end{nonfloattable}

\subsection{Cost Analysis}
\label{sec:cost-analysis}

Recorded metered generation charges were zero for the low and medium
tiers under the local and cloud-hosted Ollama configurations. The ten
complex repositories incurred about \$117. For the seven complex
repositories also run under the plain direct baseline, ADFD-Migrate cost
about 54\% as much (\$86 versus \$159). \texttt{lapack} was generated in
18 process chunks rather than one call per 3,587 files, approximately
200$\times$ fewer generation calls (\texttt{lapack} is not included in the evaluation analysis as it could not be run for all baselines/ablations). On the seven complex repositories
with repository-context direct costs, that baseline cost \$479 versus
about \$78 for ADFD-Migrate.

\subsection{Human-in-the-Loop Potential}
\label{sec:human-in-loop}

The explicit ADFD provides an artifact-level checkpoint that a developer
can inspect before target-code generation. The evaluated pipeline is
fully automatic, but the same representation supports RATIFY, REFUTE,
REVISE, and REJECT interactions in the two-way intelligibility protocol
implemented by \href{https://shraddhasurana.github.io/dhaani/}{Dhaani} (\url{https://shraddhasurana.github.io/dhaani/}).
This extends interactive DFD-based synthesis~\cite{surana2026iprog} from
code construction to automated porting.

\section{Repository-Level Results}
\label{app:repository-results}
\setcounter{table}{0}

Tables~\ref{tab:repo-low-results}, \ref{tab:repo-medium-results}, and
\ref{tab:repo-complex-results}
report the repository-level measurements underlying the aggregate results
in Section~\ref{sec:results}. Here, $\#P$ is the number of ADFD processes
and Beh. is the number of passing behavioral probes over attempted probes.

\begin{table*}[tp]
\caption{Repository-level results for the low-complexity tier.}
\label{tab:repo-low-results}
\centering
\scriptsize
\renewcommand{\arraystretch}{0.92}
\resizebox{\textwidth}{!}{%
\begin{tabular}{@{}lrrrrrrrrrrl@{}}
\toprule
\textbf{Repository} & \textbf{LoC} & \textbf{Files} &
\textbf{$\#P$} & \textbf{Chunks} &
\textbf{Syntax} & \textbf{Import} &
\textbf{Semantic} & \textbf{Beh.} & \textbf{Outcome} & \textbf{Time} & \textbf{Model} \\
\midrule
fortran2018-ex.   & 3.1k  & 75  & 19 &  7 & 100\% & 100\% & 100\% & 4/4 & 100.0\% & 25\,min & 30B \\
roots-fortran     & 5.2k  &  4  &  1 &  1 & 100\% & 100\% &  95\% & 0/7 &  97.0\% & 13\,min & 30B \\
forpy             & 13.8k &  9  &  3 &  1 & 100\% & 100\% &  95\% & 4/4 &  97.0\% & 2\,min & 30B \\
tsunami           & 3.1k  & 33  & 31 & 11 & 100\% & 100\% &  95\% & 7/7 &  97.0\% & 214\,min & 30B \\
IO-Fortran-Lib.   & 40.9k & 34  &  6 &  2 & 100\% &  70\% & 100\% & 4/4 &  96.4\% & 21\,min & 30B \\
functional-fortran & 6.6k & 27 & 10 & 4 & 100\% & 67\% & 100\% & 7/7 & 96.0\% & 13\,min & 30B \\
M\_time           & 14.7k & 58  & 37 & 13 & 100\% & 100\% &  92\% & 4/4 &  95.2\% & 51\,min & 30B \\
fortranlib        & 28.4k & 38  & 12 &  4 & 100\% &  94\% &  92\% & 2/4 &  94.5\% & 350\,min & 30B \\
CaNS              & 11.3k & 40  & 40 &  9 &  98\% &  87\% &  98\% & 1/10 &  94.2\% & 222\,min & 30B \\
fastGPT           & 2.1k  & 19  & 14 &  5 & 100\% &  53\% & 100\% & 18/18 &  94.1\% & 25\,min & 30B \\
ABAQUS            & 1.9k  &  5  &  6 &  2 & 100\% & 100\% &  88\% & 6/6 &  92.8\% & 12\,min & 30B \\
M\_strings        & 35.4k &112  & 26 &  9 & 100\% &  88\% &  93\% & 4/4 &  92.3\% & 107\,min & 30B \\
fftpack           & 4.5k  & 70  & 18 &  6 & 100\% &  90\% &  92\% & 4/8 &  92.0\% & 106\,min & 30B \\
M\_args-main      & 4.1k  & 16  & 12 &  4 & 100\% &  50\% & 100\% & 4/4 &  91.8\% & 68\,min & 30B \\
datetime-fortran  & 3.2k  & 19  & 20 &  7 & 100\% &  35\% & 100\% & 12/12 &  91.2\% & 50\,min & 30B \\
fortran-utils     & 13.1k & 96  & 14 &  5 & 100\% &  77\% &  95\% & 7/7 &  90.8\% & 99\,min & 30B \\
test-drive        & 4.3k  &  5  & 53 & 18 & 100\% &  71\% &  92\% & 8/8 &  90.3\% & 71\,min & 30B \\
bspline-fortran   & 15.8k & 19  & 23 &  8 & 100\% &  34\% &  95\% & 2/6 &  88.1\% & 30\,min & 30B \\
toml-f            & 26.0k & 94  & 17 &  6 & 100\% &  87\% &  88\% & 10/10 &  85.5\% & 110\,min & 30B \\
FKB               & 1.5k  & 17  & 20 &  6 & 100\% &  44\% &  95\% & 8/8 &  83.3\% & 36\,min & 30B \\
\midrule
\textbf{Mean} & & & & & \textbf{99.9\%} & \textbf{77.2\%} & \textbf{95.2\%} & \textbf{116/142} & \textbf{93.0\%} & \textbf{81\,min} & \\
\bottomrule
\end{tabular}}
\end{table*}

\begin{table*}[tp]
\caption{Repository-level results for the medium-complexity tier.}
\label{tab:repo-medium-results}
\centering
\scriptsize
\renewcommand{\arraystretch}{0.92}
\resizebox{\textwidth}{!}{%
\begin{tabular}{@{}lrrrrrrrrrrl@{}}
\toprule
\textbf{Repository} & \textbf{LoC} & \textbf{Files} &
\textbf{$\#P$} & \textbf{Chunks} &
\textbf{Syntax} & \textbf{Import} &
\textbf{Semantic} & \textbf{Beh.} & \textbf{Outcome} & \textbf{Time} & \textbf{Model} \\
\midrule
stdlib          & 35k  & 405 & 16  &  6 & 100\% & 100\% & 100\% & 11/11 & 98.9\% & 61\,min  & 480B \\
FOODIE          & 19k  &  44 & 14  &  5 & 100\% &  97\% & 100\% & 4/4 & 96.8\% & 31\,min  & 480B \\
arpack-ng       & 148k & 334 & 16  &  6 & 100\% &  69\% & 100\% & 7/7 & 96.3\% & 71\,min  & 480B \\
atomsk          & 86k  & 166 & 12  &  4 & 100\% &  72\% & 100\% & 4/4 & 96.1\% & 47\,min  & 480B \\
WPS             & 32k  & 123 & 16  &  6 & 100\% &  73\% & 100\% & 4/4 & 96.1\% & 53\,min  & 480B \\
packmol         & 13k  &  40 & 132 & 42 &  99\% &  74\% & 100\% & 8/8 & 96.0\% & 177\,min & 480B \\
F-A-Toolkit     & 28k  &  58 & 90  & 30 &  95\% &  92\% & 100\% & 4/4 & 95.7\% & 253\,min & 480B \\
dftd4           & 16k  &  42 & 15  &  5 &  97\% &  72\% & 100\% & 4/4 & 95.3\% & 89\,min  & 480B \\
minpack         & 10k  &  13 & 16  &  3 & 100\% &  74\% &  98\% & 9/9 & 94.9\% & 75\,min  & 480B \\
coretran        & 41k  & 116 & 17  &  6 & 100\% &  78\% &  98\% & 4/4 & 94.6\% & 59\,min  & 480B \\
Cmathtuts       & 28k  &  46 & 11  &  4 &  94\% &  69\% & 100\% & 4/4 & 94.3\% & 66\,min  & 480B \\
xtb             & 192k & 378 & 14  &  5 &  95\% &  74\% & 100\% & 4/4 & 93.9\% & 98\,min  & 480B \\
Incompact3d     & 45k  &  55 & 16  &  5 &  91\% &  60\% &  98\% & 0/5 & 91.6\% & 120\,min & 480B \\
json-fortran    & 26k  &  61 & 22  &  8 & 100\% &  62\% &  92\% & 6/6 & 90.3\% & 100\,min & 480B \\
wavelets        & 4.8k &  10 & 32  & 11 &  95\% &  88\% & 100\% & 9/9 & 89.5\% & 58\,min  & 30B  \\
SISSO           & 7.9k &  10 &  9  &  3 & 100\% &  48\% & 100\% & 4/4 & 87.0\% & 69\,min  & 30B  \\
crest           & 94k  & 178 & 13  &  5 &  98\% &  75\% &  85\% & 3/4 & 86.8\% & 91\,min  & 480B \\
neural-fortran  & 13k  & 101 & 17  &  6 & 100\% &  64\% &  95\% & 8/8 & 86.2\% & 37\,min  & 480B \\
quadpack        & 9.3k &  13 & 25  &  9 & 100\% &  29\% & 100\% & 1/12 & 81.1\% & 100\,min & 30B  \\
MPM3D-F90       & 7.8k &   9 & 44  & 15 &  99\% &  80\% & 100\% & 12/12 & 77.0\% & 106\,min & 30B  \\
\midrule
\textbf{Mean} & & & & & \textbf{98.1\%} & \textbf{72.5\%} & \textbf{98.3\%} & \textbf{110/127} & \textbf{91.9\%} & \textbf{88\,min} & \\
\bottomrule
\end{tabular}}
\end{table*}

\begin{table*}[tp]
\caption{Repository-level results for the complex tier.}
\label{tab:repo-complex-results}
\centering
\scriptsize
\renewcommand{\arraystretch}{0.92}
\resizebox{\textwidth}{!}{%
\begin{tabular}{@{}lrrrrrrrrrrl@{}}
\toprule
\textbf{Repository} & \textbf{LoC} & \textbf{Files} &
\textbf{$\#P$} & \textbf{Chunks} &
\textbf{Syntax} & \textbf{Import} &
\textbf{Semantic} & \textbf{Beh.} & \textbf{Outcome} & \textbf{Time} & \textbf{Model} \\
\midrule
lapack       & 1.56M & 3,587 & 52 & 18 & 100\% & 90\% & 100\% & 16/16 & 98.3\% & 105\,min & S4.5 \\
pymc2        & 132k  &   410 & 22 &  7 & 100\% & 80\% & 100\% & 10/10 & 96.7\% & 98\,min & S4.5 \\
ccpp-physics & 325k  &   240 & 45 & 15 & 100\% & 78\% & 99.9\% & 7/11 & 96.6\% & 135\,min & S4.5 \\
Nek5000      & 281k  &   327 & 33 & 12 & 100\% & 90\% & 98\% & 13/13 & 96.5\% & 232\,min & S4.5 \\
petsc        & 29k   &   200 & 25 &  9 & 100\% & 80\% & 100\% & 8/8 & 96.2\% & 115\,min & S4.5 \\
hdf5         & 94k   &   158 & 85 & 29 & 100\% & 77\% & 100\% & 10/10 & 95.7\% & 107\,min & S4.5 \\
elmerfem     & 1.07M & 2,211 & 35 & 12 & 100\% & 88\% & 100\% & 8/8 & 95.7\% & 162\,min & S4.5 \\
fpm          & 48k   &   217 & 44 & 15 & 100\% & 59\% & 100\% & 8/16 & 94.5\% & 100\,min & S4.5 \\
openfast     & 505k  &   330 & 34 &  9 & 100\% & 66\% & 100\% & 11/11 & 94.2\% & 108\,min & S4.5 \\
cp2k         & 1.33M & 1,325 & 36 &  5 & 100\% & 90\% & 92\% & 10/10 & 93.1\% & 65\,min & S4.5 \\
\midrule
\textbf{Mean} & & & & & \textbf{100.0\%} & \textbf{79.8\%} & \textbf{99.0\%} & \textbf{101/113} & \textbf{95.7\%} & \textbf{123\,min} & \\
\bottomrule
\end{tabular}}
\end{table*}


\section{Reproducibility Details}
\label{app:reproducibility}
\setcounter{table}{0}

\subsection{Representative Behavioral Outputs}

Table~\ref{tab:behavioral-output-values} gives representative
Fortran-oracle comparisons. The full set of probe inputs, tolerances, and
outputs is included in the replication package.

\begin{nonfloattable}
\caption{Representative outputs from the ADFD-Migrate executable Fortran-vs-Python
harness. \emph{Max. abs.} is largest absolute difference between the
Fortran oracle and generated Python.}
\label{tab:behavioral-output-values}
\centering
\small
\setlength{\tabcolsep}{3pt}
\renewcommand{\arraystretch}{1.15}
\begin{tabular}{@{}>{\raggedright\arraybackslash}p{0.27\columnwidth}%
                  >{\raggedright\arraybackslash}p{0.23\columnwidth}%
                  >{\raggedright\arraybackslash}p{0.23\columnwidth}%
                  r@{}}
\toprule
\textbf{Repository /} & \textbf{Input} & \textbf{Common} &
\textbf{Max.} \\
\textbf{routine} & & \textbf{output} & \textbf{abs.} \\
\midrule
tsunami /\newline \texttt{num\_tiles}
  & \texttt{n=6}
  & \texttt{[3, 2]} & 0 \\
lapack /\newline \texttt{dcopy}
  & \texttt{x=[5,-1,}\newline\texttt{0,2.5,7]}
  & \texttt{[5,-1,}\newline\texttt{0,2.5,7]} & 0 \\
minpack /\newline \texttt{enorm}
  & \texttt{x=[1e-20,}\newline\texttt{3,4,1e19]}
  & \texttt{1.0e19} & 0 \\
cp2k /\newline \texttt{kahan\_}\newline\texttt{dot\_product}
  & \texttt{a=[1,2,3],}\newline\texttt{b=[4,5,6]}
  & \texttt{32.0} & 0 \\
arpack-ng /\newline \texttt{compute\_}\newline\texttt{residual\_}\newline\texttt{norm}
  & \texttt{r=[-3,}\newline\texttt{-4,0,0]}
  & \texttt{5.0} & 0 \\
FKB /\newline \texttt{gaussian}
  & \texttt{x=[-2,}\newline\texttt{-0.5,0,1.25]}
  & \texttt{[0.01832,}\newline\texttt{0.77880,}\newline\texttt{1.0, 0.20961]} & 0 \\
stdlib /\newline \texttt{gauss\_}\newline\texttt{legendre\_}\newline\texttt{points}
  & \texttt{n=3, [-1,1]}
  & \texttt{[-0.77460,}\newline\texttt{0, 0.77460]}
  & $4.4{\times}10^{-16}$ \\
\bottomrule
\end{tabular}
\end{nonfloattable}

\subsection{Computational Environment}

Experiments ran on a MacBook Pro (Mac16,5) with an Apple M4 Max
(16-core CPU, integrated GPU, and 48\,GB unified memory) under
macOS~26.5.1 (Darwin~25.5.0). Qwen-3-Coder 30B inference used local
Ollama acceleration; Qwen-3-Coder 480B and Anthropic inference were
remote. Configuration manifests record models, sampling parameters,
token use, system information, and stage timings.


\section*{Acknowledgements}
During part of this work, AS was a visiting Professorial Fellow at UNSW,
and a Visiting Professor at the Centre for Health Informatics at
Macquarie University. He is a member of the Anuradha and Prashant
Palakurthi Centre for AI Research (APPCAIR) at BITS Pilani. The authors
would like to thank Santonu Sarkar and Soumyadip Bandyopadhyay for their comments on the paper.

\section*{Declaration of generative AI and AI-assisted technologies in the manuscript preparation process}
During preparation of this manuscript, the authors used AI-assisted
drafting tools to help organize empirical material, integrate tables,
and improve wording. The authors reviewed and edited the content and
take full responsibility for the manuscript.


\bibliographystyle{cas-model2-names}
\bibliography{references}



\end{document}